%% file: korinna18.tex
\documentclass[epj,nopacs]{svjour}
\usepackage[english]{babel}
\usepackage{amsmath}
\usepackage{amssymb}
\usepackage{units}
\usepackage{graphicx}
\usepackage{cite}

\newcommand{\gsim}{~{\buildrel > \over {_\sim}}~}
\newcommand{\pt}{p_\perp}

\newcommand{\df}{\Delta \phi}
\renewcommand{\d}{{\rm d}}

\newcommand{\mean}[1]{\langle #1 \rangle}
\newcommand{\der}[2][]{\frac{\text{d}#1}{\text{d}#2}}

\begin{document}

\headnote{\hfill \small \rm CERN-PH-TH/2008-067}

\title{A Monte Carlo Model for 'Jet Quenching'}
\author{Korinna Zapp\inst{1,2}\and
Gunnar Ingelman\inst{3}\and
Johan Rathsman\inst{3}\and
Johanna Stachel\inst{2}\and
Urs Achim Wiedemann\inst{1}}

\institute{Physics Department, Theory Unit, CERN, 
CH-1211 Gen\`eve 23, Switzerland\and
Physikalisches Institut, Universit\"at Heidelberg, Philosophenweg 12,
D-69120 Heidelberg, Germany\and
High Energy Physics, Uppsala University, Box 535, S-75121 Uppsala, Sweden}

\mail{zapp@physi.uni-heidelberg.de}

\abstract{ 
We have developed the Monte Carlo simulation program \textsc{Jewel}~1.0 (Jet Evolution
With Energy Loss), which
interfaces a
perturbative final state parton shower with medium effects occurring in
ultra-relativistic
heavy ion collisions. This is done by comparing for each jet fragment the
probability
of further perturbative splitting with the density-dependent probability of scattering
with the medium. A simple hadronisation mechanism
is
included. In the absence of medium effects, we validate \textsc{Jewel} against
a set of
benchmark jet measurements. For elastic interactions with the medium, we
characterise not
only the medium-induced modification of the jet, but also the jet-induced
modification
of the medium. Our main physics result is the observation that collisional and
radiative
medium modifications lead to characteristic differences in the jet fragmentation
pattern,
which persist above a soft background cut. We argue that this should allow to
disentangle
collisional and radiative parton energy loss mechanisms by measuring the 
$n$-jet fraction or a class  of jet shape observables.
}

\maketitle

\section{Introduction}
\label{sec1}

In ultra-relativistic heavy ion collisions, the produced QCD matter reduces
significantly the
energy of high transverse momentum partons. Experiments at the Relativistic
Heavy Ion
Collider RHIC give strong support
to this picture~\cite{Adcox:2004mh,Back:2004je,Arsene:2004fa,Adams:2005dq}. Most
of the experimental evidence comes from studying the leading hadronic fragments
of the parent partons via single inclusive hadron spectra and jet-like particle
correlations. Since the 
medium-induced suppression of these spectra persists unattenuated up to the
highest
transverse momenta accessed experimentally at RHIC, it  is likely to play a
dominant
role also in the much wider transverse momentum 
range soon to be explored in heavy ion collisions at the CERN Large Hadron Collider 
LHC~\cite{Carminati:2004fp,Alessandro:2006yt,D'Enterria:2007xr}.

On general grounds, the energy lost by the leading fragment in a parton shower
must
manifest itself in the associated particle yield. So, parton energy loss is
expected to result
in a medium modification of the {\it entire} parton fragmentation pattern
(`jet quenching')
and not only in the suppression of leading hadrons. The corresponding study 
of medium-modified  jets beyond their leading fragments is of great interest for
several
reasons, in particular: i)~At the LHC a larger fraction of the entire
medium-modified jet
fragmentation pattern will become accessible above background. ii)~Studying the
distribution of subleading fragments is likely to discriminate between different
microscopic
mechanisms conjectured to underly jet quenching, thereby helping to characterise
more
precisely the properties of matter tested by jet quenching. iii)~Modelling the
distribution
of subleading jet fragments is essential for any operational procedure aiming at
disentangling jets from background or characterising the jet-induced modification
of
the background. Such reasons motivate the development of tools which account
dynamically
for the interaction between jet and medium, and which model medium-modified
jets on the level of multi-particle final states. 

The interactions of a partonic jet component with some target 
(typically, a quark or a gluon) in the medium can be elastic or inelastic. 
In the elastic case, the projectile parton
can lose energy
only by transferring recoil momentum to the target ('collisional parton energy
loss' ~\cite{Bjorken:1982tu,Thoma:1990fm,Braaten:1991we,Djordjevic:2006tw,Adil:2006ei,Zakharov:2007pj,Peigne:2008nd,Domdey:2008gp,Zapp:2005kt}).
In the inelastic case, the dominant source of energy degradation of the partonic
projectile
is the medium-induced break-up of the projectile via gluon radiation ('radiative
parton energy
loss' ~\cite{Gyulassy:1993hr,Baier:1996sk,Zakharov:1997uu,Wiedemann:2000za,Gyulassy:2000er,Wang:2001if}), resulting
in two or more projectile components of lower energy.

The Monte Carlo (MC) technique provides a powerful tool for the simulation of
multi-particle
final states. In the absence of medium-effects, the perturbative dynamics of
parton fragmentation
maps with known accuracy onto a probabilistic iteration of parton splittings,
and the Monte Carlo
technique is widely used for the simulation of final state parton showers in the
vacuum~\cite{Sjostrand:2006za,Marchesini:1991ch,Gleisberg:2003xi}.
In the presence of medium effects, we expect that a parton shower can contribute
to understanding `jet quenching', 
in particular for the following reasons:

\begin{enumerate}
\item {\it Baseline for the study of medium-modified multi-particle states.}\\
The standard experimental procedure of characterising medium-modifications of
jet-like
multi-particle final states is to compare to an experimental baseline in which
medium
effects are absent. A realistic modelling should be able to parallel this
experimental
procedure. 
\item {\it Exact implementation of energy and momentum conservation.}\\
Multi-particle final states are known to be sensitive to trigger biases and they
are constrained
to share the energy and momentum of a parent parton. For the description of this
constrained dynamics, exact energy-momentum conservation is likely to be
important.
It may be numerically more important than the treatment of interference effects,
on
which analytical calculations of radiative parton energy loss tend to focus.
\item {\it Testing a wide range of microscopic mechanisms for the interaction
between
projectile and target.}\\
A Monte Carlo tool should allow the user to change easily between
different descriptions of the medium and of the interaction
between the
medium and the partonic projectile. This is important to constrain the
microscopic mechanism
underlying parton energy loss and to better identify the specific properties
of the medium tested by jet quenching.
\item{\it Interface with experiment.}\\
 The modelling of realistic multi-hadron final states is clearly beneficial for
 comparing theory and data, including detailed detector acceptance, resolution 
 and response. 
\end{enumerate}

We note that the application of Monte Carlo techniques for the modelling
of medium-effects leaves open conceptual
issues in the current description of jet quenching. In particular,
medium-induced gluon radiation
is expected to arise from an interference phenomenon whose mapping onto a 
probabilistic description of multi-gluon emission is not known. Also, while the
iteration of collisional
mechanisms may be probabilistic, its interface with a standard parton shower 
involves modelling assumptions. In the present work, we have bypassed these
fundamental issues in order to arrive at a tool which realises the practically
important points listed above. 

The working name of this Monte Carlo code is \textsc{Jewel},
standing for Jet Evolution With Energy Loss. In the absence of a medium, the \textsc{Jewel}
parton shower is
ordered in virtuality, and strict
angular ordering is enforced by phase space constraints. The medium is modeled
as a collection of  scattering centres whose cross sections and distribution in
phase
space can, in principle, be chosen freely. To decide whether a parton splits
as in
the vacuum or whether 
it interacts with a target parton in the medium, the code compares the mean free path
between
scattering centres with the mean lifetime of the virtual parton. Within this
framework,
\textsc{Jewel} aims at realising the objectives listed above.

One approach currently used to connect parton energy loss calculations
to measurable quantities is based on energy loss probabilities 
('quenching weights'~\cite{Baier:2001yt,Salgado:2003gb}), defined by 
the probabilistic iteration of single medium-induced gluon emissions.
Alternatively, the interpretation of medium-induced gluon radiation as 
a modification of parton splitting functions has motivated attempts to evolve 
partons via medium-modified DGLAP evolution equations.
Both approaches lead to comparable results for leading hadronic
fragments~\cite{Armesto:2007dt}. Also, collisional energy loss mechanisms have
been used recently to define quenching weights and medium-modified DGLAP 
evolutions~\cite{Domdey:2008gp}. We note that these approaches 
are largely limited to leading hadron spectra, or special classes of two-particle 
correlation functions~\cite{Majumder:2004br}, mainly since they treat subleading
fragments with kinematic approximations. 
Recently, there have been many efforts to go beyond leading hadron spectra 
within the framework of the models discussed above. This included works on 
jet broadening~\cite{Salgado:2003rv}, the angular dependence of 
two-particle correlations~\cite{Polosa:2006hb},
the effect of directed momentum transfer
(a.k.a. flow) on the jet fragmentation~\cite{Armesto:2004pt,Armesto:2004vz},
the modification of jet multiplicity distributions~\cite{Salgado:2003rv,Borghini:2005em} and 
jet hadrochemistry~\cite{Sapeta:2007ad}. 
We expect that Monte Carlo models of jet quenching will greatly advance this field of study, mainly because they can implement the four important features listed above. 

In recent years, several Monte Carlo models implementing
parton energy loss have been developed already. In particular, the code
\textsc{Hijing}~\cite{Wang:1991hta} simulates complete nucleus-nucleus collisions and provides a simplified model for radiative energy loss derived from analytical calculations. There are also codes, which simulate jet quenching without modeling the entire nucleus-nucleus collision, such as
 \textsc{Pqm}~\cite{Dainese:2004te}, which is a Monte Carlo implementation of the BDMPS quenching weights, and \textsc{Pyquen}~\cite{Lokhtin:2005px}, which
modifies the standard \textsc{Pythia~6.2} jet events essentially by reducing 
the energy of partons in the shower and by adding additional gluons to that shower according to distributions motivated by parton energy loss calculations.
There are also two models of medium modified parton showers, namely \textsc{Q-Pythia}~\cite{Armesto:2008qh} with modified splitting functions including induced gluon radiation in the BDMPS model and a model where the virtuality of partons in the shower is increased due to interactions with the medium thus stimulating additional gluon emissions~\cite{Renk:2008pp}. All these approaches go beyond leading hadron spectra. In developing \textsc{Jewel}, our main focus was to arrive at a code which allows to study in detail the dynamics relating the evolution of the parton shower to the microscopic modeling of the medium interactions.  In particular, we are interested in specifying dynamically the physically interesting (but model-dependent) relation between parton energy loss, $\pt$-broadening, recoil momentum, change in jet multiplicity, and other characteristic features of parton energy loss.

This paper describes the physics encoded in \textsc{Jewel} version 1.0, and 
demonstrates its use for the
study of jet quenching phenomena on the level of multi-hadron final states. 
The paper is organised as follows: In section~\ref{sec2}, we discuss the Monte
Carlo
final state parton shower and its interface with a hadronisation mechanism in
the
absence of medium effects. To arrive at a manageable inclusion of medium-effects
on the hadron level, the hadronisation model used by \textsc{Jewel} is less
sophisticated than
what is implemented in state of the art Monte Carlo event generators. We discuss these
differences and show that \textsc{Jewel} provides an adequate description of many jet
observables,
including event shapes, $n$-jet fractions and intra-jet multiplicity
distributions.
In section~\ref{sec3}, we discuss how scattering in the plasma is included in
\textsc{Jewel}.
Most of our discussion will focus on the case of elastic scatterings.
For this case, we test in section~\ref{sec3} a well-studied theoretical
baseline, namely the
in-medium propagation of a parton in the absence of parton splitting. 
Section~\ref{sec4} addresses the question 
to what extent the medium-modifications included in
\textsc{Jewel} affect
experimentally accessible jet observables in heavy ion collisions. 
\textsc{Jewel}, as any newly developed Monte Carlo event generator, is a work in
progress. In the conclusions, we summarise what we have learned from the present 
exploratory study and highlight some directions for future work.

\section{The Monte Carlo Model in the absence of medium effects}
\label{sec2}

In this section, we introduce the baseline on top of which medium-effects are
included and show that it basically reproduces the observed QCD radiation
physics of jets in the vacuum. The evolution variable of a parton shower is not
unique. We have decided to use the virtuality $Q^2$. As explained in more detail 
in section~\ref{sec3}, this  has the
advantage that the evolution variable traces the lifetime $1/Q$ of the
virtual
states, which facilitates the embedding of the parton shower in 
the spatiotemporal geometry of a medium. 
We interface this parton shower with a hadronisation scheme which implements
the idea that colour neutralisation occurs locally during hadronisation. However,
the
scheme invoked here is less sophisticated than the hadronisation prescriptions
used
in modern event generators, in particular in that it does not require knowledge
about the
event-specific colour flow in the parton shower. This is a technical
simplification,
which - in contrast to standard treatments - allows for a straightforward
extension of the
hadronisation mechanism in the presence of a medium.

\subsection{Final state parton shower in the absence of medium effects}
\label{sec2a}
\subsubsection{Parton evolution}
We want to describe the evolution of a parton of initial energy $E$, 
produced in a hard scattering process. This parton fragments into a
multi-parton final state. In the absence of a medium, the \textsc{Jewel}
parton shower is closely related to the
mass-ordered shower in the \textsc{Pythia}~6.4 event
generator~\cite{Sjostrand:2006za}.
The kinematics of
each $a \to b+c$ parton branching is given in terms of the virtuality of the
parent parton
and the momentum fraction $z$ carried by one of its daughters. The probability
that
no splitting occurs between an initial and final virtuality $Q_i$ and $Q_f$,
respectively,
is described by the Sudakov factor
\begin{eqnarray}
\label{eq_sudavac}
\lefteqn{ S_{\text a}(Q_{\rm i}^2,Q_{\text f}^2)
   = } \\
& & \exp \left[ - \int \limits_{Q_{\text f}^2}^{Q_{\text i}^2} \!
     \frac{{\text d} Q'^2}{Q'^2} \!\!\int \limits_{z_-(Q'^2,E)}^{z_+(Q'^2,E)}
\!\!\!\!\!\!\!
     {\text d} z \,
     \frac{\alpha_{\text s}(z(1-z)Q'^2)}{2\pi}\, \sum_{\text b,c}
     \hat P_{{\text a}\to {\text bc}}(z) \right]\, . \nonumber
\end{eqnarray}
Here, $\hat P_{\text{a}\to\text{bc}}(z)$ are the standard LO parton splitting
functions for quarks and gluons ($a,b,c \in \{q,g\}$). In order to regularise
the integral one has to define an infrared cut-off scale below which splittings
are considered to be not resolvable. In practice, we require a minimal virtual
mass $Q_0/2$ for the daughters. Given this cut-off, the condition $k_\perp^2 \approx z(1-z)Q^2
> \Lambda_\text{QCD}^2$ translates directly into the allowed $z$ range
\begin{equation}
 z_\pm(Q^2,E) = \frac{1}{2} \pm \frac{1}{2}
                 \sqrt{\left( 1 - \frac{Q_0^2+4 \Lambda_\text{QCD}^2}{Q^2}\right) \left( 1 -
                 \frac{Q^2}{E^2}\right) }\, .
\end{equation}
To avoid the Landau pole in equation~(\ref{eq_sudavac}), we increase the nominal
cut on $k_\perp$ by \unit[10]{\%}. These are the same prescriptions as used in the mass-ordered \textsc{Pythia} cascade.
With the no-splitting probability equation~(\ref{eq_sudavac}) the probability
density $\Sigma_{\text a}(Q_{\text i}^2,Q^2)$ for a splitting to happen at
virtuality $Q^2$ is given by
\begin{equation}
 \Sigma_{\text a}(Q_{\text i}^2,Q^2) 
      = \frac{{\text d} S_{\text a}(Q_{\text i}^2,Q^2)}{{\text d} (\ln Q^2)}
      = S_{\text a}(Q_{\text i}^2,Q^2) \sum_{\text b,c}
        W_{{\text a}\to {\text bc}}(Q^2)\, ,
        \label{2.3}
\end{equation}
where
\begin{equation}
 W_{{\text a}\to{\text bc}}(Q^2)
       =  \int \limits_{z_-(Q^2,E)}^{z_+(Q^2,E)}\!\!\!\!\!
     {\text d} z \, \frac{\alpha_{\text s}(z(1-z)Q^2)}{2\pi}
     \hat P_{{\text a}\to{\text bc}}(z)
\end{equation}
is the differential probability for the splitting $a \to b+c$ at
$Q^2$, and the Sudakov form factor $S_{\text a}(Q_{\text i}^2,Q^2)$ denotes the 
probability for evolving from $Q_{\text i}^2$ to $Q^2$ without splitting.

We determine the virtuality $Q_{\text a}^2$ of the parent parton according to
the probability density $\Sigma_{\text a}(E^2,Q_{\text a}^2)$. Consistent with
the
probability distributions $W_{{\text a}\to{\text bc}}$, we select the type of
parton splitting, which occurred for parton $a$. 
The momentum fraction $z$ of the splitting is then chosen within the
kinematically allowed
range\\ $z \in \left[ z_-(Q_\text{a}^2,E), z_+(Q_\text{a}^2,E) \right]$ for the
decay of a parton of virtuality $Q_\text{a}$,
requiring that the virtualities of both partons are larger than the hadronisation scale parameter $Q_0/2$.  
Subsequently, the virtual masses of the two daughter partons are determined with the help
of equation (\ref{2.3}), subject to three constraints: The virtualities $Q_b$,
$Q_c$ of the
daughters are required to be smaller than their energy $z\, E$ or $(1-z)E$,
respectively,
and they must be larger than the hadronisation scale $Q_0/2$. In addition, the
virtual masses of the daughters satisfy the constraint $Q_b^2 + Q_c^2 < Q_a^2$.
The branching $a \to b + c$ is finally accepted if the momentum fraction $z$ chosen initially lies within the kinematically
allowed range
for daughters of 'mass' $Q_b$ and $Q_c$, respectively. Otherwise, new values for $Q_b$ and $Q_c$ are chosen, in simulations with angular ordering it may also be necessary to reject the $z$ value and try with a new one.
If the branching is
accepted, the
full four-momentum is reconstructed for both daughters, assuming an azimuthally
isotropic
decay of the parent. The procedure is then iterated
for the daughter partons, based on their assigned virtualities. 
Parton evolution is stopped for daughter partons
which do not
split above the scale $Q_0^2$ and these partons are declared to be on mass shell.
Angular ordering is enforced by allowing only splittings with decreasing
emission angle. The coupling $\alpha_\text{s}$ is running at one loop. 
This shower is essentially the 'global' 'constrained' evolution which is one of the alternatives
of the \textsc{Pythia} event generator.

\subsubsection{Hadronisation mechanism}

In order to have a hadronisation mechanism which is flexible enough to also be useable for the complex parton state in heavy ion collisions, we have developed a modified application of the string hadronisation approach.
The hadronisation prescription assumes maximal colour correlation between
partons close in momentum space. 
The code identifies first the parton with the highest energy in the event
(alternatively, the highest $p_\perp$ may be used in hadron collisions.)
If this parton is a gluon, it is split into a collinear quark -
antiquark pair with the energy sharing given by $\hat
P_{\text{g}\to\text{q\= q}}$. The more energetic of the two is then the
endpoint of the first string. 
In the next step it is connected to the  closest parton in momentum space
(with the only exception that a quark-antiquark pair from a
single gluon splitting is not allowed to recombine into a colour singlet).
In case the closest parton is a gluon the procedure is iterated until either
another (anti)quark is connected or there is no parton in the same hemisphere
left.
In the latter case a suitable endpoint is generated by adding the required
quark or antiquark with momentum in the beam direction to the event. 
The whole procedure is repeated until all partons are connected in strings. The
strings are then hadronised using the Lund string fragmentation  \cite{lund} routine 
of \textsc{Pythia}~6.4, with default values of hadronisation
parameters. We checked that the strings are sufficiently massive for
this string fragmentation routine to apply. 

This approach is inspired by the fragmentation of jets in hadronic
collisions where the additional endpoints can be thought of as being part of the
proton remnants, to which the jet is connected by colour flow. The hadrons
associated
with this additional parton endpoint tend to go along the beam direction so that
they
are well separated from the jet. As long as the jet structure is analysed in a 
restricted rapidity range of approximately $|\Delta \eta| \le 1$ around the
rapidity of the parent parton, the resulting  dependence of the model on this 
endpoint is negligible. There is the possibility to set a maximum
invariant mass
of neighboring partons in a string, that could be used to tune the routine to
data.
However, we did not attempt to fine-tune this simplified hadronisation routine. 

This hadronisation mechanism has the advantage of being very flexible. It
can be applied to any number of jets in an event and, more importantly, it 
can be applied also to jets in a nuclear environment. In the latter case, it may 
be desirable to hadronise only
some high energy part of the event. This is also possible, since this
hadronisation routine can also handle systems that are not globally
colour neutral.

For $e^+e^-$ collisions,  beam remnants do not exist and the two hemispheres 
are colour connected and do not hadronise separately.
We therefore implemented a variation of the string finding
algorithm, which takes into account these additional constraints on colour flow.
For the observables discussed in this paper, we did not identify any significant
difference
between the two hadronisation mechanisms (MC results not shown). To simplify our 
presentation, we therefore decided to simulate all observables with the same
hadronisation routine described above, irrespective of whether they are for
$e^+e^-$  or hadronic collisions. 

We checked that the hadronic
observables are essentially independent of the cut-off scale $Q_0$, at which
the perturbative evolution is interfaced with string hadronisation. The
default choice is $Q_0=\unit[1]{GeV}$.

\subsection{Comparison to data}
\label{sec2b}
%
%
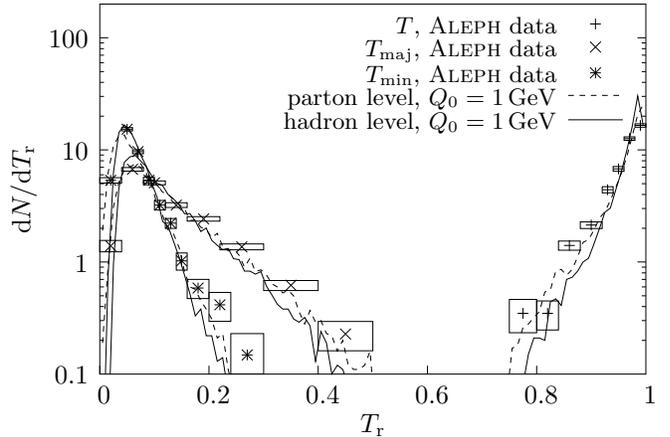
\begin{figure}[t]
\centering
\input{1-thrustvac-lqcd300.pstex_t}
\caption{The thrust, thrust major and thrust minor ($T_\text{r}=(T,T_\text{maj},T_\text{min})$) distributions for
$\sqrt{s}=\unit[200]{GeV}$ $e^+e^-\to q\, \bar{q}\to X$
collisions. Data of the ALEPH Collaboration~\cite{Heister:2003aj} 
are compared to simulations of \textsc{Jewel}: i)~parton level after parton shower evolved down to $Q_0 = \unit[1]{GeV}$, ii)~hadron level after parton shower evolution followed by hadronisation ($Q_0 = \unit[1]{GeV}$).}
\label{fig1}
\end{figure}
%

\begin{figure*}[t]
\centering
\input{2-njetvac-lqcd300.pstex_t}
\caption{The jet rates as a function of jet resolution scale $y_{\rm cut}$ in $\sqrt{s}=\unit[200]{GeV}$ $e^+e^-\to q\, \bar{q}\to X$
collisions.
Left hand side: Simulation of \textsc{Jewel} with and without hadronisation for evolution down to $Q_0=\unit[1]{GeV}$. Right hand side: Data of the \textsc{Aleph}
collaboration~\cite{Heister:2003aj}
compared to simulations of \textsc{Jewel} with hadronisation.}
\label{fig2}
\end{figure*}
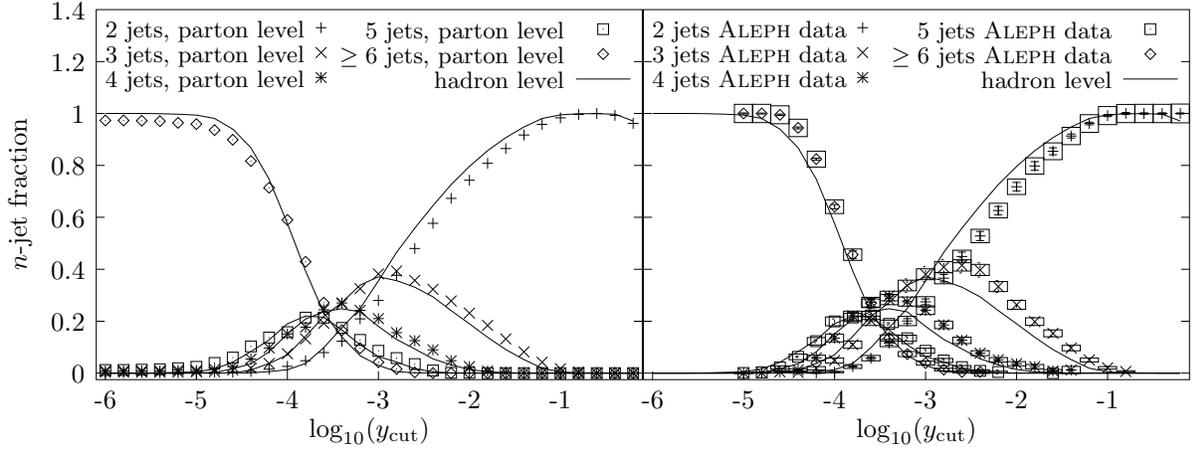
%

In this subsection, we compare the final hadronic state from \textsc{Jewel} to
data on jets measured in $\sqrt{s}=\unit[200]{GeV}$ $e^+e^-$ collisions  at
LEP by the ALEPH collaboration~\cite{Heister:2003aj}. To select the process
$e^+e^- \to q\, \bar{q} \to X$ at $\sqrt{s}=\unit[200]{GeV}$, these ALEPH data 
were taken with a veto on initial state radiation. They have been 
compared already~\cite{Heister:2003aj}
very favourably to standard event generators such as \textsc{Pythia} and
\textsc{Herwig}. The purpose of the present data comparison is to validate
the \textsc{Jewel} parton shower  in the absence of medium effects
against a set of benchmark data on jets measured at LEP, before studying the
extension of the code to medium effects. 

To account for a 2-jet event, \textsc{Jewel} evolves each parent parton
separately. For $e^+e^-$ data, the code determines the initial virtuality of 
one of the two parent quarks according to the probability distribution 
$\Sigma_{\text a}(E^2, Q^2)$, given in equation (\ref{2.3}). Then, there 
are two options: Either, the virtuality of the second parent
quark is fixed to be the same, thus conserving energy and momentum at the hard
vertex, or we choose it independently according to the same distribution
(\ref{2.3}). Since the differences observed between these schemes turned out to 
be negligible, we present here only results for the latter case. 

We focus on three classes of tests. First, we test \textsc{Jewel} on the level
of the overall energy
flow, which is characterised by event shapes. Then we study an
observable
which is particularly sensitive to the parton fragmentation pattern. Finally, we
turn to
a comparison of  inclusive charged hadron distributions inside a jet.
Hadronisation
will be seen to play a more important role for the latter observables.

In figure~\ref{fig1}, we compare data to simulations of three event shape
observables,
namely thrust $T$, thrust major $T_{\rm maj}$ and thrust minor $T_{\rm min}$.
For these
quantities, one sums over the three-momenta $\vec{p}_i$ of all final state
particles.
According to the definition of thrust,
\begin{equation}
	T \equiv {\rm max}_{\vec{n}_T} 
	\frac{\sum_i \vert \vec{p}_i\cdot \vec{n}_T\vert}{\sum_i \vert
\vec{p}_i\vert}\, ,
	\label{2.5}
\end{equation}
a 2-jet event is pencil-like if $T=1$, that is if all particles are aligned
parallel or antiparallel
to a thrust axis $\vec{n}_T$. The event is spherical if $T=1/2$.  Once the
thrust axis
$\vec{n}_T$ is known, one can determine the direction $\vec{n}$ orthogonal to
$\vec{n}_T$,
along which the momentum flow is maximal. Thrust major is defined  as the
projection
of all particle momenta on this direction $\vec{n}$,
\begin{equation}
	T_{\rm maj} \equiv {\rm max}_{\vec{n}_T\cdot \vec{n}=0} 
	\frac{\sum_i \vert \vec{p}_i\cdot \vec{n}\vert}{\sum_i \vert
\vec{p}_i\vert}\, .
	\label{2.6}
\end{equation}
Thrust minor sums up the components $\vec{p}_{ix}$ of the final particle
momenta $\vec{p}_{i}$, which are  orthogonal to the plane defined by
$\vec{n}$
and $\vec{n}_T$,
\begin{equation}
	T_{\rm min} \equiv 
	\frac{\sum_i \vert \vec{p}_{ix}\vert}{\sum_i \vert \vec{p}_i\vert}\, .
	\label{2.7}
\end{equation}
We note that $T$, $T_{\rm maj}$ and $T_{\rm min}$ are perturbatively calculable, infrared-safe quantities and therefore particularly suitable for testing our parton shower implementation.

As seen in figure~\ref{fig1}, the final state parton shower provides a reasonable description
of these jet event shapes over most of the measured range. We recall that
\textsc{Jewel}
does not contain a matching of the parton cascade  to exact matrix elements,
which
could improve the QCD modelling of large angle radiation. This may be the reason
why the simulation gives fewer events with large 
$T_{\rm  maj}$ and $T_{\rm min}$. Also, we observe small deviations between
data and simulation for very small values of $T_{\rm  maj}\, ,T_{\rm min} <
0.05$ at the parton level.
This region is known \cite{Korchemsky:1995zm,Dokshitzer:1997ew} to be very sensitive to non-perturbative effects
and hence cannot be reliably described by a leading order parton cascade as 
illustrated by the dashed curve. Consistent with this
general statement, we find that for $T_{\rm  maj}\, ,T_{\rm min} < 0.05$, 
the inclusion of the hadronisation prescription leads to some improvement of 
the data comparison. 

We have compared \textsc{Jewel} to measurements of a wider class of
event shape observables, in particular to oblateness, sphericity, planarity,
aplanarity
and total jet broadening. The comparison to these event shape observables are
of similar or better quality than the comparisons shown in
figure~\ref{fig1}. With
these studies, we have established that \textsc{Jewel} accounts
for global features of jet energy flow with an accuracy which is sufficient to 
characterise (sufficiently large) medium effects on top of it. 

In comparison to jet event shapes, there are measurements which are more
sensitive to
the discrete and stochastic nature of partonic processes underlying 
the QCD jet fragmentation. One such measurement is the so-called
$n$-jet-fraction.
Its definition is based on the Durham clustering algorithm \cite{Catani:1991hj}. 
For each pair of final state particles, one defines a distance 
\begin{equation}
y_{ij} = 2 {\rm min}(E_i^2,E_j^2) (1-\cos \theta_{ij}) / E_{\rm cm}^2\, .
\label{2.8}
\end{equation}
The pair of particles with smallest $y_{ij}$ is then replaced by a
pseudo-particle,
whose energy and momentum are the sums of its daughters. The clustering
procedure is repeated until all $y_{ij}$ exceed a given threshold $y_{\rm cut}$.
The number of clusters separated by a distance larger than $y_{\rm cut}$ is
defined to be the number $n$ of jets.  Thus, as one decreases the resolution 
scale $y_{\rm cut}$, one becomes sensitive to finer and finer details of the
discrete QCD radiation pattern.

Figure~\ref{fig2} shows simulation results for the $n$-jet fraction. We find
that
for $\log_{10}(y_{\rm cut}) \gsim -3$, the jet resolution scale is sufficiently
coarse such that
hadronisation plays a negligible role in the $n$-jet-fraction. It is only for
smaller
scales, that the hadronic late stage of QCD fragmentation affects the number
of jets identified by the Durham clustering algorithm. The 3-jet fraction is somewhat too small at large $\log_{10}(y_{\rm cut})$ while the 2-jet fraction is correspondingly too large, which is again due to the missing 3-jet matrix element.
We find that the $n$-jet fractions vary mildly with the size of the strong coupling on the partonic as well as on the hadronic level. Our choice of $\Lambda_\text{QCD}$ leads to a reasonably good description of the data.

\begin{figure}[t]
\centering
\input{3-xivac-lqcd300.pstex_t}
\caption{The inclusive distribution $\d N_{\rm ch}/\d \xi$, $\xi = \ln
\left[ E_\text{jet}/p_\text{hadron} \right]$ of charged hadrons in $e^+e^-\to
q \bar q \to X$ events at $\sqrt{s}=\unit[200]{GeV}$.
Data of the ALEPH Collaboration~\cite{Heister:2003aj} 
are compared to simulations of \textsc{Jewel}: i)~parton level after parton shower evolved down to $Q_0 = \unit[1]{GeV}$, ii)~hadron level after parton shower evolution to $Q_0 = \unit[1]{GeV}$ followed by hadronisation.}
\label{fig3}
\end{figure}
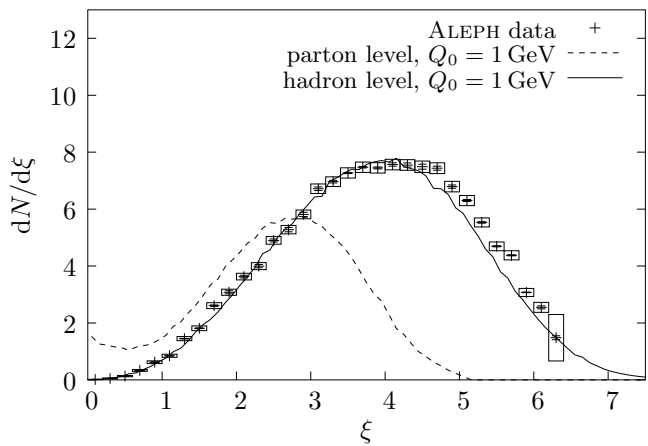
%
In contrast to the measurements discussed so far, the modelling of single
inclusive
intra-jet hadron distributions and multi-hadron correlations requires detailed
knowledge
about the hadronisation mechanism. This is seen for instance in
figure~\ref{fig3}, where
we compare results of our simulation to data of the inclusive distribution $\d N_{\rm ch}/\d \xi$, ( $\xi =
\ln \left[ E_\text{jet}/p_\text{hadron} \right]$) of charged hadrons in
$e^+ e^-\to q \bar q \to X$ events at $\sqrt{s}=\unit[200]{GeV}$. Irrespective of
the scale $Q_0$ down to which
the parton
cascade is evolved prior to hadronisation, there is a marked difference between
the hadronic
and the partonic distribution. In particular, the partonic distribution is very
'hard', i.e.
the yield of high momentum (small $\xi<1$) partons exceeds the observed
hadronic yield by far. 
Since high momentum partons are correlated in colour to softer partons, it is
likely that any colour
neutralisation
mechanism, i.e.\  hadronisation, which takes into account colour flow, will fill the momentum space between these
partonic partners. As a consequence, hadronisation is expected to soften the 
distribution for $\xi < 1$ considerable. This is seen for the string
hadronisation mechanism, which accounts for the inclusive hadron distribution $\d N_{\rm ch}/\d \xi$ (figure~\ref{fig3}) to a level better than a few \%.

We also made simulations with the independent string hadronisation mechanism~\cite{Finkelstein:1972qm,Krzywicki:1973qh,Field:1977fa}, which does not invoke any 
colour correlation between partons and instead hadronises each parton in 
the final state separately. This mechanism was found to depend strongly on $Q_0$ and tends to produce too many soft hadrons.

We finally note that in the presence of a
high-multi\-plicity environment, novel hadronisation mechanisms may play a role. For
instance, hadron formation may occur via recombination of 
partons~\cite{Molnar:2003ff,Fries:2003vb,Greco:2003xt,Hwa:2002tu,Fries:2003kq}. 
Studying the hadrochemistry of jets~\cite{Sapeta:2007ad,Liu:2008zb} is likely to help 
characterising such novel hadronisation mechanisms, but a study of this is beyond the scope of this paper.

Within these uncertainties, which are mainly related to the modelling of hadronisation,
we have established that \textsc{Jewel} provides a reliable baseline for the
characterisation of jet quenching phenomena.

\section{`Jet Quenching' in the Monte Carlo Model}
\label{sec3}

A Monte Carlo model of jet quenching requires a picture of how a final state
parton shower is embedded in the space-time geometry of a nuclear 
collision, and how it interacts with the medium. The question where the  
parton splittings occur in space is likely to affect the resulting medium 
modifications strongly, since it determines which partonic components 
interact for how long with the medium. However, very little is known about the 
spatiotemporal evolution of parton showers in the vacuum or in the medium.
Inverting the standard logic, one may even view the medium as a probe of the
spatiotemporal evolution of the jet, since the medium interferes with parton 
fragmentation on length scales comparable to time-dilated hadronisation times. 
This highlights that understanding the medium-modification of the jet
relies on understanding how the medium interacts with it, and 
understanding properties of the medium relies on how they are reflected in 
the medium-modification of jets. Progress on this mutual dependence requires 
a dynamical description of both the jet {\it and} the medium.

On general grounds, one expects that a parton of virtuality $Q$ branches on a
time
scale $1/\vert Q\vert$ in its rest frame, which translates to a Lorentz
time-dilated length
scale in the rest frame of the nuclear matter. Our model of the spatiotemporal
distribution
of parton branchings within a nuclear environment will be based on this relation
between
time scales and virtuality. While different parton showers use slightly
different evolution variables,
we have chosen virtuality (consistent with the \textsc{Pythia}~6.4 mass-ordered
shower),
since it makes the contact with
a spatiotemporal picture relatively easy. 

\subsection{Modelling the interactions with the medium}
\label{sec3a}

We regard the medium as a collection of partons acting as scattering centres. For the case of
elastic
interactions between the jet and the medium, each scattering centre displays to
the partonic
projectiles an elastic $2\to 2$ scattering cross section $\d\sigma/\d t$. The
form of this cross section, as well as the momentum distribution, the mass
and the density of scattering centres can be specified freely in \textsc{Jewel}.
For the studies presented
in this paper, we focus on a specific model, in which the density and momentum
distributions of scattering centres are that of a gas of massive quarks and gluons
of
temperature $T$. The masses of the scattering centres are
fixed to $m_\text{scatt}=\mu_\text{D}(T)/\sqrt{2}$, where $\mu_\text{D}(T)$ is
the
thermal Debye mass. It is approximated by $\mu_\text{D}=3T$ in the simulation. This model is minimal in the sense that  the medium is
characterised fully by a single parameter, the temperature $T$. 

To specify the spatiotemporal structure of the parton shower, we start from
the
estimate that the parton shower evolves down to components of virtuality $Q_f$ 
on a time scale $\sim 1/Q_f$. For a parton of energy $E$ and mass $Q_f$, this will
be
time-dilated in the rest frame of the medium to $\sim E/Q_f^2$. 
If the parton originated from the branching of some parton of virtuality $Q_i$,
then
the parton of virtuality $Q_f$ existed  for a duration of approximately
\begin{equation}
	\tau = \frac{E}{Q_f^2} - \frac{E}{Q_i^2}\, .
	\label{3.2}
\end{equation}
For the parent parton, which initialised the parton shower, the lifetime is
$\tau =  \frac{E}{Q_i^2}$. In the case of a medium of  constant density $n$, 
the probability that no scattering occurs during this time is given by
\begin{equation}
	S_{\rm no\,  scatt}(\tau) = \exp\left[  - \sigma_{\rm elas}\, n\, \tau \beta
\right]\, .
	\label{3.3}
\end{equation}
Here, $\sigma_{\rm elas}$ is the total elastic scattering cross section and $\beta$ the projectile velocity. For the evaluation of the total scattering cross section in equation~(\ref{3.3}) the scattering centres are assumed to be at rest, for the simulation of the scattering process they are assigned a thermal momentum.
If the density of scattering centres varies along the trajectory of the parton,
one would
replace this expression by an integral over the parton trajectory between
initial
and final times, $\exp\left[  - \int_{\tau_i}^{\tau_f} \sigma_{\rm elas}(\xi)\,
n(\xi) \d \xi \right]$.
Here, we have included the possibility that the elastic cross section changes
during
evolution, for instance, since its infrared regulation depends on properties of
the medium, or $\alpha_\text{s}$ has a temperature dependence.

For the differential elastic scattering cross sections, we choose
\begin{equation}
	\frac{{\rm d}\sigma}{{\rm d}\vert t\vert} = \frac{\pi\alpha_s^2}{s^2} C_R
		\frac{s^2+u^2}{ |t|^2}\Bigg\vert_{\rm regularised}\, .
		\label{3.1}
\end{equation}
This is the leading $t$-channel exchange term for quark-quark ($C_R = 4/9$),
quark-gluon ($C_R = 1$) and gluon-gluon ($C_R = 9/4$) scattering, respectively. 
The regularisation of equation~(\ref{3.1}) and the running of the coupling will
be specified below. Instead of equation~(\ref{3.1}), one could include the full LO $2\to 2$
parton cross sections~\cite{Combridge:1978kx}. These contain, for instance,
also terms singular in the Mandelstam variable $u$. For a scattering in the medium, this corresponds
to processes in which almost all the projectile energy is transferred to the
scattering
partner. Such processes may be viewed not as an energy degradation of the
leading parton but as an exchange of the roles of projectile parton and target parton.
For the sake of simplicity, the present study does not aim at including and
characterising
such more detailed features of elastic interactions between projectile and
target.

In the Monte Carlo model studied here, we know the virtuality of the partons,
which
emerge from parton splittings or from elastic scatterings: 
After a splitting, the virtuality is determined via equation~(\ref{2.3}), after
a scattering, it remains unchanged. This prescription is consistent with the
dominance
of small angle $t$-channel scatterings, which do not open phase space that could
be
used to reduce the fast parton's virtuality.
Given this virtuality, we determine the lifetime $\tau$ of the parton
according to equation~(\ref{3.2}). With the probability $S_{\rm no\,  scatt}(\tau)$, this
parton will
not scatter, and the code simulates the next splitting
as in the
vacuum case. With the probability $\left[1-S_{\rm no\,  scatt}(\tau)\right]$,
the parton
scatters at some time $\tau'<\tau$. In this case, the parton exchanges momentum
with a scattering centre according to the differential cross section
equation~(\ref{3.1}).
Then, the  code continues to propagate the scattered projectile parton, by
checking
whether there will be further parton scatterings before splitting. In this
procedure,
angular ordering is enforced for consecutive splittings, but it is reset at each
scattering,
consistent with the assumption that scattering destroys the interference pattern
of
parton radiation.
On-shell partons are allowed to scatter until they leave the medium. It
 is thus assumed that partons cannot hadronise inside the medium.
The code has an option to propagate also the recoiling
target partons with subsequent splittings and scatterings. To simplify the
presentation
and discussion, we do not explore this option in the present work.

The above prescription for embedding a parton shower in a nuclear environment
includes inevitably model-de\-pen\-dent assumptions.  In this exploratory work, 
we do not embark on their systematic study, but we would like to recall some
of the major sources of such model dependencies:

First, as mentioned already, the spatiotemporal structure of the parton shower
is based
on assumptions, which are difficult to constrain. For instance, even if the
lifetime
$\tau$ of a virtual state is given parametrically by (\ref{3.2}), one may
distribute this
lifetime with an exponential decay law with half time (\ref{3.2}), or one may
try to
improve the lifetime estimate e.g.\ by accounting more accurately for the energy
of the daughter
partons or by replacing it with other assumptions. 

Second, the choice of the cross section (\ref{3.1})
invokes the assumption that the interaction between the jet and the medium can
be
treated perturbatively. This assumption is problematic, since data at RHIC
provide
some evidence of anomalously strong coupling. In addition, there are
quantitative
uncertainties arising from the regularisation of the cross section (\ref{3.1}).

Third, the
model introduced here does not yet include a mechanism of radiative energy
loss.
Since this may be the main source of energy degradation, we have included an
option
to enhance the vacuum splitting functions by a factor $(1 + f_{\rm med})$,
\begin{equation}
	\hat{P}_{a\to bc}(z) \longrightarrow (1 + f_{\rm med})\, \hat{P}_{a\to
bc}(z)\, ,
	\label{3.4}
\end{equation}
as long as the splitting occurs in the medium. This prescription has been argued
to mimic characteristics of radiative energy loss~\cite{Borghini:2005em}. 
In a probabilistically iterated parton shower,  a more satisfactory treatment of
radiative energy loss may be to include partonic inelastic 
$2\to 3$ scattering cross sections on the same footing as the elastic $2 \to 2$
cross sections included here.  We did not implement this
option in the present study, since its discussion involves significant
additional
conceptual and technical issues, which - in our view - deserve a separate
study.

Fourth, there is the question of how to interface a medium-modified parton
shower
with a hadronisation model.
In the present study, the final partonic state is interfaced
with the hadronisation model discussed in section~\ref{sec2} after all
interactions with the medium.
So, the present version of \textsc{Jewel} is based
on the assumption that the hadronisation of a jet fragment is unaffected by
medium
effects, but this assumption could be modified in future studies to explore for
instance
changes in jet hadrochemistry~\cite{Sapeta:2007ad}.

The present paper does not aim at fully studying or resolving any of these 
issues. But we expect that \textsc{Jewel} provides a tool for their 
systematic exploration in subsequent works.

\subsection{A baseline of medium-effects: collisional energy loss without parton
branching}
\label{sec3b}
The case of a single, high energy parton losing energy via multiple elastic
interactions
in a spatially extended target has been studied extensively in the literature
~\cite{Bjorken:1982tu,Thoma:1990fm,Braaten:1991we,Djordjevic:2006tw,Adil:2006ei,Zakharov:2007pj,Peigne:2008nd,Domdey:2008gp,Zapp:2005kt}.
This problem can be studied in \textsc{Jewel} by switching off the option of
parton splitting, and specifying the (model-dependent) density of scattering
centres and
elastic scattering cross sections. The MC then simulates the propagation of an on-shell parton (massless or massive) which undergoes scattering in a medium. The scattering probability is in this case determined only by the path length inside the medium, and the density and cross section of scattering centres. Here, we only consider the collisional energy loss of massless partons and we leave the case of heavy flavours to a dedicated study.

\begin{figure}[t]
\centering
\input{4-dedx-lqcd300.pstex_t}
\caption{The average parton energy loss $\d E/\d x$ of a quark of energy $E$,
undergoing multiple elastic collisions over a path length $L = \unit[1]{fm}$ 
in a thermal medium of temperature $T$. Elastic 
collisions are described by the infra-red regulated partonic cross sections
of equation (\ref{3.5}) (case~I) and equation (\ref{3.6}) (case~II).
}\label{fig4}
\end{figure}
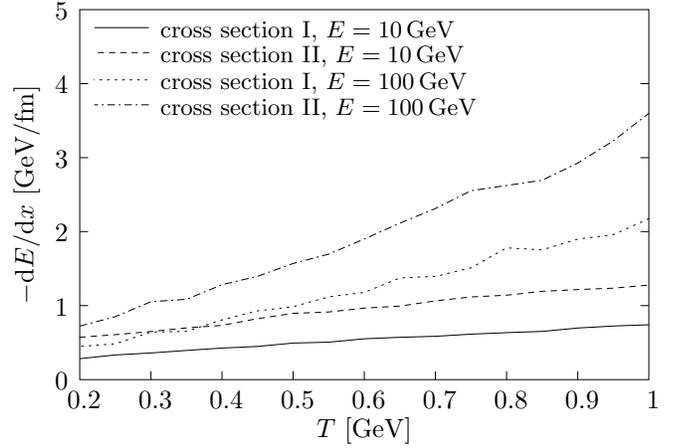
%
As discussed in section~\ref{sec3a}, we define the
density, mass and momentum distribution of scattering centres in terms of a
thermal
distribution. The same temperature is the model parameter entering the
regularisation
of elastic scattering cross sections. To make contact with the previous studies,
we
haven chosen two different regularisation schemes for the elastic scattering
cross section,
namely case~I
\begin{equation}
	\sigma^{\rm elas} = \int \limits_0^{\vert t_{\rm max}\vert} d\vert
t\vert\,
	\frac{\pi\, \alpha_s^2(\vert t\vert + \mu_D^2)}{s^2}
	C_{R}\, \frac{s^2 + (s-\vert t\vert)^2}{\left(\vert t\vert + \mu_D^2
\right)^2}\, ,
	\label{3.5}
\end{equation}
which is the default, and case~II
\begin{equation}
	\sigma^{\rm elas} = \int \limits_{\mu_D^2}^{\vert t_{\rm max}\vert} d\vert
t\vert\,
	\frac{\pi\, \alpha_s^2(\vert t\vert )}{s^2}
	C_{R}\, \frac{s^2 + (s-\vert t\vert)^2}{\vert t\vert ^2}\, .
	\label{3.6}
\end{equation}
The kinematic boundary is $\vert t_{\rm max}\vert = 3
\sqrt{2} E T$. The total cross sections (\ref{3.5}) and
(\ref{3.6}) differ only
by their
regularisation. 
In figure~\ref{fig4}, we have calculated the resulting average energy loss for
an in-medium path length of $L = \unit[1]{fm}$ in a medium of temperature
$T$. In general, the average collisional energy
loss increases with increasing temperature and projectile energy.
Quantitatively, the
models I and II show differences of approximately a factor 2 in $\d E/\d
x$ for a \unit[10]{GeV} parton. The differences decrease slowly with increasing
projectile energy. Cross section~II leads to a larger energy loss, 
as may be expected since there is minimum momentum transfer.

%
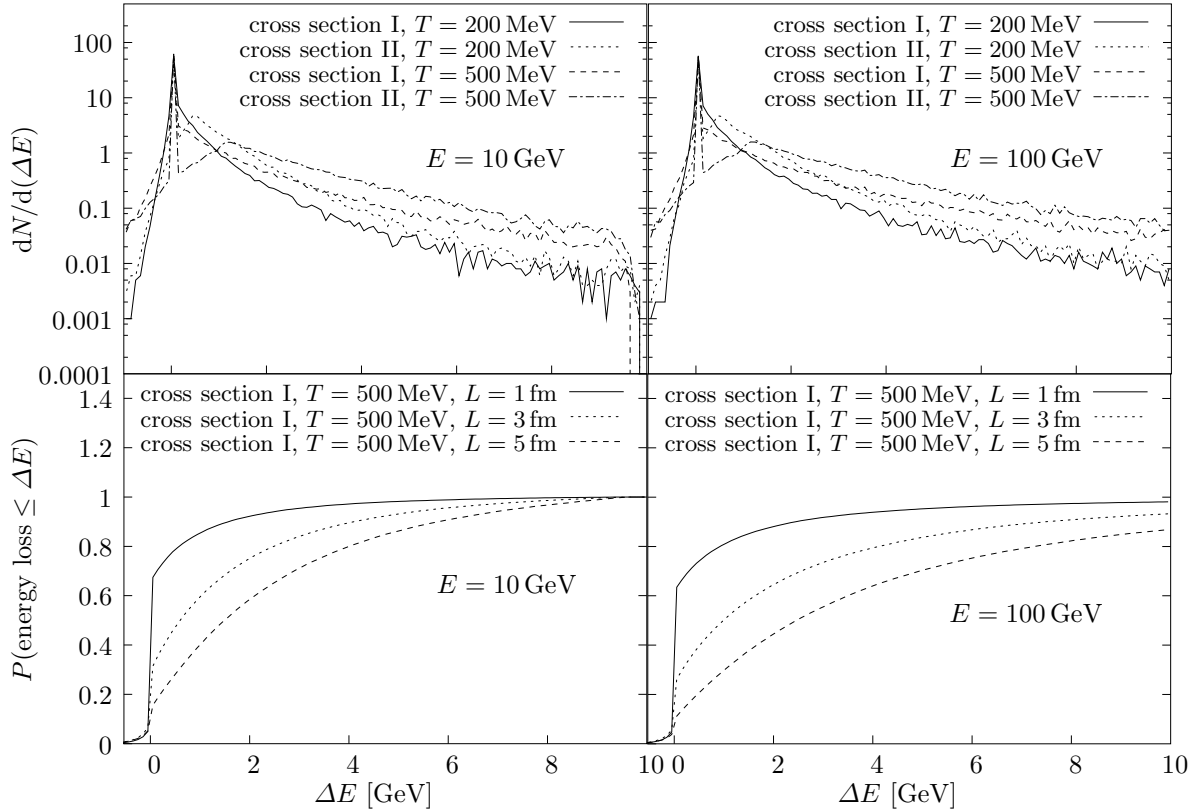
\begin{figure*}[t]
\centering
\input{5-deltae-lqcd300.pstex_t}
\caption{Distribution of energy loss $\Delta E$ for different parameter
choices (top panels) and the probability for an energy loss smaller than 
$\Delta E$ after passage of a medium of length $L$ (bottom panels).}
\label{fig5}
\end{figure*}
%

We note that in calculations based on finite temperature field theory,
the medium specifies a preferred Lorentz frame and the effect of collisional
energy loss is in general not fully described by an elastic scattering cross 
section of the form $\d\sigma/\d t$. This complicates analytical comparisons
of our model with recent studies of collisional energy loss~\cite{Bjorken:1982tu,Thoma:1990fm,Braaten:1991we,Djordjevic:2006tw,Adil:2006ei,Zakharov:2007pj,Peigne:2008nd,Domdey:2008gp,Zapp:2005kt}. 
On the other hand, to the best of our knowledge, no characteristic medium-induced
deviations from the form $\d\sigma/\d t$ have been explored so far, and a comparison
of our calculations for $\d E/\d x$ on the numerical level seem meaningful. 
We find that the temperature dependence of $\d E/\d x$ shown in figure ~\ref{fig4},
is consistent with the dependences reported previously. The differences between 
case~I and case~II are representative of the typical factor 2 uncertainties between 
different model studies, though some recent studies lead to slightly larger values
of  $\d E/\d x$ than those shown in figure~\ref{fig4}, see e.g.~\cite{Peigne:2008nd}.
Here, we do not enter the current discussion of whether an improved understanding
of the scale dependence of elastic interactions may allow to narrow these uncertainties. 
Rather, we take equation (\ref{3.5}) as an example of an elastic cross section, whose
strength and $t$-dependence may be changed in future simulations of 
\textsc{Jewel}.
We finally note that in a Monte Carlo simulation, the average energy loss does not 
grow linearly with the in-medium path length. This is because energy and
momentum are conserved exactly so that subsequent collisions occur at decreasing
centre of mass energies, which introduces a mild non-linearity. 

%
\begin{figure*}[t]
\centering
\input{6-bckgrnd-lqcd300.pstex_t}
\caption{
Transverse momentum relative to beam axis and angle with
respect to jet axis of
recoiling scattering centres as compared to the undisturbed medium for
different temperatures, with and without splitting of the projectile (cross
section~I, $E_\text{jet}=\unit[100]{GeV}$, $L=\unit[1]{fm}$). The
jet is at mid-rapidity ($\eta = 0$). Hadronisation is not included but may affect these distributions significantly (see text for further discussion).}
\label{fig6}
\end{figure*}
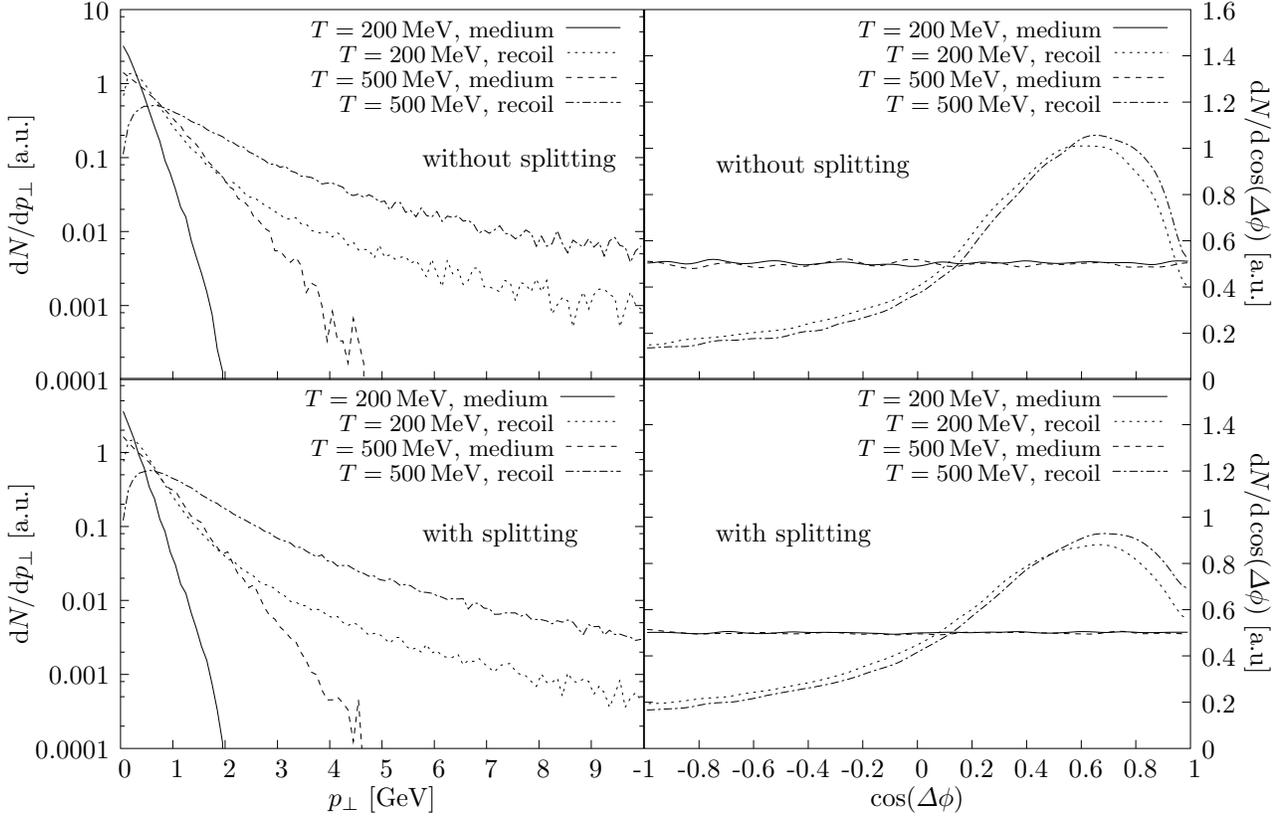

For single inclusive spectra, which fall steeply in transverse momentum, the dependence of parton energy loss is not determined by the average energy loss but by the least energy loss which a significant fraction of the projectiles suffer. For this reason, it has become the standard in studies of collisional energy loss to study $\d N/\d \Delta E$ of partons suffering an energy loss $\Delta E$. Our result for this distribution is shown in figure~\ref{fig5} and is seen to peak sharply for negligible energy loss. This reflects the fact that elastic interactions are dominated by small angle scattering, for which the longitudinal component of the momentum transfer between projectile and target becomes negligible in the high energy limit. 

We note that the temperature dependence of collisional parton energy loss models shows several generic features. Qualitatively, these may be understood by observing that the strength of the average collisional energy loss is essentially governed by the product of the density of scattering centers times the cross section, which each of these scattering centers displays. The density rises proportional to $T^3$ with increasing temperature. The cross section, however, is dominated by the lowest typical momentum transfer, which is governed by the Debye screening mass, so  parametrically $\sigma \propto 1/m_D^2 \propto 1/T^2$. For increasing temperature, the strong increase in density is largely balanced by a
reduction of the effective cross section per scattering center, and the net effect is an average energy loss $\Delta E$ which shows a mild, approximately linear increase with $T$, as seen in Fig.~\ref{fig4}. For the same reason, increasing the Debye screening mass by increasing $T$, one reduces the relative yield of scatterings at small $t$ strongly, while the rate of particles propagating without interaction through a finite size medium is affected only mildely. As a consequence, the peak in the distribution $\d N/\d\Delta E$ at $\Delta E = 0$ depends on details of the infrared regularization of the elastic
 cross section, but tends to become more pronounced with increasing temperature. The mean free path changes only mildly with temperature but is
relatively large (\unit[1.7]{fm} for a \unit[100]{GeV} quark at $T=\unit[500]
{MeV}$).

Another characteristic property of $t$-channel dominated elastic differential
cross sections of the form equation~(\ref{3.1}) is their weak $\sqrt{s}$-dependence
in the high energy limit. This leads to a remarkably weak dependence of
$\d N/\d \Delta E$ on projectile energy $E$, as seen in figure~\ref{fig5}. We note,
however, that the $\d N/\d \Delta E$ distribution extends up to the limit
$\Delta E = E$.  As a consequence, the average energy loss
$-\frac{\d E}{\d x} = \int_0^E \d(\Delta E)\, \frac{\d N}{\d \Delta E}\, \Delta E
\Big/  \int_0^E \d(\Delta E)\, \frac{\d N}{\d \Delta E}$ differs significantly
for $E = \unit[10]{GeV}$ and $E = \unit[100]{GeV}$, since for larger jet
energy $E$, the integrand $ \frac{\d N}{\d \Delta E}\, \Delta E$ has contributions
for larger values of $\Delta E$. As seen in figure~\ref{fig4}, there is a marked $E$-dependence
of the average energy loss, while there is a very weak $E$-dependence of 
$ \frac{\d N}{\d \Delta E}$ for small values of $\Delta E$. This illustrates the well-known 
fact (for more details, see e.g. Ref.~\cite{Baier:2001yt}) that the value of the average energy loss is dominated by rare events with 
large $\Delta E$, and thus $-\d E/\d x$ does not characterise adequately the
'typical' energy loss suffered by most events entering $\frac{\d N}{\d \Delta E}$.

Upon proper normalisation, the distribution $\d N/\d \Delta E$ can be turned
into a probability distribution for a high energy quark to lose less than a 
specific amount of energy $\Delta E$. This quantity is plotted in the lower 
panel of figure ~\ref{fig5}. One sees that for an in-medium path-length of
$L=\unit[1]{fm}$,
almost 65 \% of all projectile partons do not lose any energy in our
models,
even if the temperature is taken to be \unit[500]{MeV}. For \unit[3]{fm}
in-medium
path length, there are still typically 25 \% of all projectiles which
emerge
unscathed. This indicates that elastic interactions alone are unlikely to 
account for a large fraction of the suppression of single inclusive hadron
spectra. 

\subsection{Characterising the Recoiling Medium}

%
\begin{figure}[t]
\centering
\includegraphics[scale=0.475]{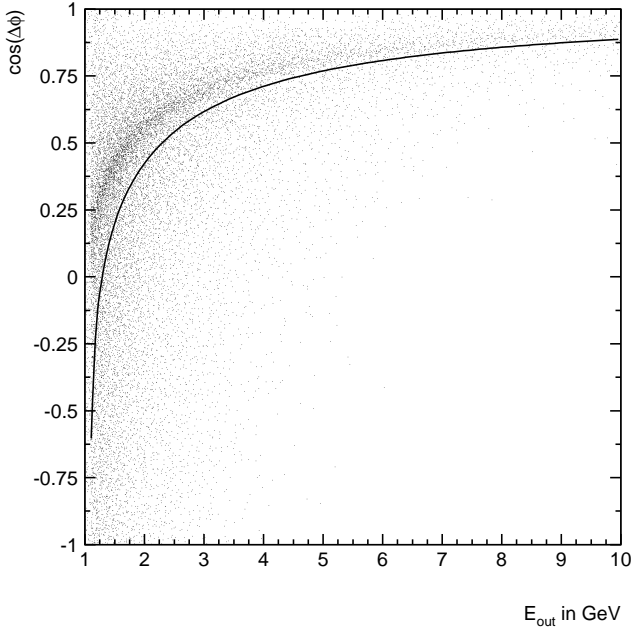}
\caption{Correlation of angle and momentum of recoiling scattering
centres in the Monte Carlo simulation ($T=\unit[500]{MeV}$, without splitting, $L =\unit[1]{fm}$,
$E = \unit[100]{GeV}$) and
analytical estimate (equation~(\ref{eq_cosdf-p-cor}) for mean $E_\text{in}$)}
\label{fig7}
\end{figure}

The energy lost by a jet is redistributed amongst the target components.
If one could characterise the amount of energy recoiling in the medium, this
would
be an unambiguous determination of collisional effects. (We note that also
$2\to 3$ and $2\to n$ processes can contribute to such recoil effects.) More
generally,
characterising the recoil of the medium may provide a means to disentangle
different mechanisms of parton energy loss. In this section, we characterise
the momentum distribution of target partons, recoiling against a projectile 
parton which undergoes elastic $2\to 2$ processes in a thermal medium.

In figure~\ref{fig6} the momentum
and angular distribution of recoiling scattering centres is compared to the
undisturbed thermal medium. We investigate the cases, that i) 
an on-shell projectile is propagated without splitting (upper panel of
figure~\ref{fig6}) or
that ii) the splitting of the projectile is included (lower panel of
figure~\ref{fig6}).
The results are very similar for both cases, indicating that the recoil is only
weakly
sensitive to the projectile energy.

The scattering centres in the medium show initially an
exponentially falling $\pt$-spectrum and an isotropic momentum distribution,
characteristic for a thermal distribution. After interaction with the
projectile,
the $\pt$-spectrum of the recoil follows a power law at intermediate and high
$\pt$.
The high-$\pt$ tail is more pronounced at higher temperatures.
Using the kinematics of $2\to 2$ scattering, one estimates 
\begin{equation}
E_\text{out} = \frac{|t| + 2 E_\text{in}^2}{2 E_\text{in}} \qquad \Rightarrow
\qquad
\der[\sigma]{E_\text{out}} = 2 E_\text{in} \der[\sigma]{|t|}\, ,
\end{equation}
where $E_\text{in}$ and $E_\text{out}$ are the energies of the incoming and
outgoing scattering centre, respectively, and the average over the angle between
the initial partons was taken. We note that despite this power law, the
yield of recoil particles lies mainly at relatively soft momentum, $\pt < \unit[1-2]{GeV}$. 

The recoil moves predominantly in the jet direction with a maximum at
$\df \sim 0.8$ nearly independent of the temperature. The results shown
here are for
cross section~I; with option II the result is qualitatively the same. There is
a moderate change in the shape of the distribution, but the position of the
maximum remains unchanged.

Most of the features of the angular correlation can be understood from a
simplified analytical model where the scattering centre is assumed to be at rest
and the cross section is approximated by $\d \sigma/\d |t| \propto |t|^{-2}$
(very similar results are obtained with $\d \sigma/\d |t| \propto (|t|+\mu_\text{D}^2)^{-2}$). 
In this approximation $\cos(\Delta\phi)$ can only be positive.
Given the shape of the distribution the position of the maximum can be
approximated by the expectation value
\begin{equation}
\mean{\cos(\Delta\phi)} \simeq \left(\sqrt{3}-1\right) +
\frac{m_\text{scatt}}{E_\text{p}} \left(2\sqrt{3}-3\right)\, ,
\end{equation}
which is practically constant for large projectile energies $E_\text{p}$. 
It has, in particular, only a weak temperature dependence through the mass
$m_\text{scatt}$ of the scattering centre. However, the position of the 
maximum is determined by the regularisation of the cross section. We
therefore investigate the dependence on the infrared regulator by treating it
as an independent variable. After expansion around $\mu_\text{D}^2=2
m_\text{scatt}^2$
one gets for the shift of the maximum position
\begin{equation}
\Delta \mean{\cos(\Delta\phi)} \simeq \left[ \sqrt{3}
-1-\frac{1}{\sqrt{3}} \right] \left( \frac{\mu_\text{D}^2 -
2m_\text{scatt}^2}{2m_\text{scatt}^2} \right)
\end{equation}
This finding is in agreement with the results from the full Monte Carlo
 simulation.

There is a strong correlation between the scattering angle and the momentum
transfer, such that the most energetic partons are closer to the jet axis. If one
allows the incoming scattering centre to carry momentum but averages over 
its direction, then the analytical model discussed above leads to an angular
dependence of the outgoing scattering centre with respect to the incoming projectile,
\begin{equation}
\cos(\Delta\phi) = \frac{E_\text{out}\, (E_\text{p} +
E_\text{in}) - E_\text{p}E_\text{in} - E_\text{in}^2}{p_\text{p}\,
\sqrt{E_\text{out}^2-m_\text{scatt}^2}}\, .
\label{eq_cosdf-p-cor}
\end{equation}
In figure \ref{fig7} this result is compared to the Monte Carlo
simulation. The overall behaviour is similar, but the Monte Carlo result is
distributed wider due to the spread in energy and angle of the incoming
scattering centres.

At face value, figure~\ref{fig6} indicates that a jet can be accompanied by
additional multiplicity which has its maximum separated from the jet axis
by a characteristic finite angle $\Delta\phi$, and whose yield is expected
to die out quickly with increasing transverse momentum. Similarly,
figure~\ref{fig7}
indicates that scatterings with large energy transfer lead to a relatively large
projectile
energy loss of the leading parton, but that the recoil parton is very
close to the jet axis so that the energy may stay within the jet cone. Typical
momentum transfers, on the other hand, tend to scatter the recoil to a
characteristic and relatively large angle. We note, however, that whether the
partonic distributions of figure~\ref{fig6} and \ref{fig7} will or will
not
change
substantially upon hadronisation may depend on details of the hadronisation 
mechanism. In particular, if the partons  entering the peak of the $\cos
(\Delta\phi)$-
distribution are connected by colour strings with the leading jet fragments at
$\cos (\Delta\phi)\sim 1$, then the hadronic distribution will peak much closer to
$\cos (\Delta\phi) = 1$ than the partonic one shown in figure~\ref{fig6}. On the
other hand, multiple gluon exchanges with the target break the colour flow
between hard projectile fragments and target recoils. If such a mechanism 
is invoked, then the hadronic distribution is likely to follow the partonic one 
shown in figure~\ref{fig6}. Here, we limit our discussion to the kinematic 
constraints of the underlying partonic distributions in figure~\ref{fig6} and 
~\ref{fig7}, but we do not explore the model dependence
associated with the hadronisation of these distributions. 

\section{Medium-modifications of jet measurements} 
\label{sec4} 
In this section, we study to what extent the jet measurements discussed in
section~\ref{sec2}
are sensitive to the medium effects introduced in section~\ref{sec3}.
A medium does not only affect jet fragmentation. It also results in a high
'background'
multiplicity of soft particles. This complicates the characterisation of jets
and their
medium modifications. To establish which jet measurements are experimentally
feasible
despite this background, one would ideally like to embed the parton shower
simulated
by \textsc{Jewel} into full event simulation of a heavy ion collision. This
could be done but it lies beyond the scope
of the present paper. In this section, we compare analyses of the full
simulated
jets with analyses, for which we only include hadrons above a background cut of $E_{\rm cut}=\unit[2]{GeV}$. This comparison may indicate the extent to which different  jet
medium modifications remain visible above the hadronic background of the heavy ion collision.
%
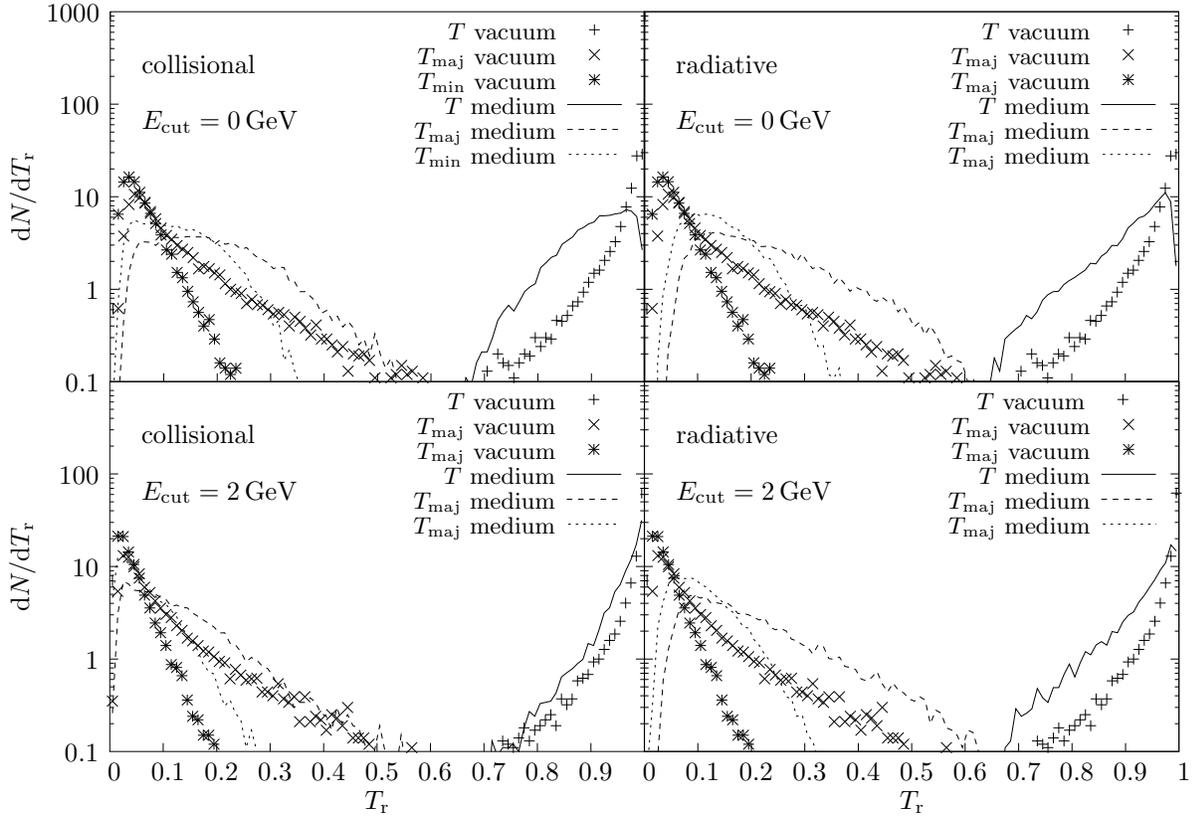
\begin{figure*}[t]
\centering
\input{9-thrustmed-lqcd300.pstex_t}
\caption{Thrust, thrust major and thrust minor ($T_\text{r}=(T,T_\text{maj},T_\text{min})$) for a single
\unit[100]{GeV} jet.
The \textsc{Jewel} parton shower in the vacuum is compared to
two scenarios including medium-induced parton energy loss. Left hand side:
Collisional
energy loss for a medium of $T=\unit[500]{MeV}$ and in-medium path length
$L=\unit[5]{fm}$ (the recoil is hadronised with the medium).
Right hand side: Radiative energy loss for $f_{\rm med} = 3$ and
$L=\unit[5]{fm}$. For
the
plots in the lower panel, only hadrons with energy 
above $E_{\rm cut} = \unit[2]{GeV}$ are included.}
\label{fig8}
\end{figure*}
%

For $e^+e^-$-collisions, thrust, thrust major and thrust minor are of particular
interest
for testing QCD radiation physics, since they are infrared safe and thus
perturbatively
calculable quantities. Applying the definitions (\ref{2.5}), (\ref{2.6}) and
(\ref{2.7}) to
a single jet or to the jet activity above background, the feature of infra-red
safety
is lost. But $T$, $T_{\rm maj}$ and $T_{\rm min}$ will still provide a
characterisation
of the global energy flow in a jet. Here, we ask to what extent these quantities
may
provide a useful characterisation of jet medium modifications. Figure~\ref{fig8}
shows
the result of simulations of the \textsc{Jewel} parton shower. If collisional
energy loss is
included, then the jet is expected to broaden.
This is clearly seen in the
broadening
of  $T$, $T_{\rm maj}$ and $T_{\rm min}$ in the upper left panel of
figure~\ref{fig8}.
However, the kinematics of elastic $2\to 2$ scatterings dictates that the more
energetic projectiles are deflected by smaller angles. We checked that the
collisional broadening
observed in figure~\ref{fig8} is due to recoil. If the recoiling scattering
centres are removed from the final state and do not hadronise, the thrust
distributions with medium are practically indistinguishable from the vacuum
results. The recoiling scattering centres have mostly relatively low
momenta so that the medium-induced broadening is
much reduced and becomes small if only hadrons of energy above $E_{\rm
cut}=\unit[2]{GeV}$ are included in the analysis (see lower left panel of
figure~\ref{fig8}).
In contrast to collisional energy loss, single partonic components of
medium-induced
additional radiation have a higher probability to carry a significant energy
fraction of
the initial projectile energy. So, on general grounds, one expects that the
medium-induced
broadening of the distributions  in thrust, thrust major and thrust minor will
persist even
if soft hadrons of energy $E_\text{h} < E_{\rm cut} = \unit[2]{GeV}$ are dropped
from the analysis.
This is clearly seen in figure~\ref{fig8}. We note that the value of $f_{\rm med}$
entering these simulations is a priori a free model parameter. We estimated, however, 
that the choice $f_{\rm med} = 3$ lies within the range of parameters consistent
with a nuclear modification factor $R_{\rm Pb\, Pb} \approx 0.2$ in central
lead-lead collisions. 

%
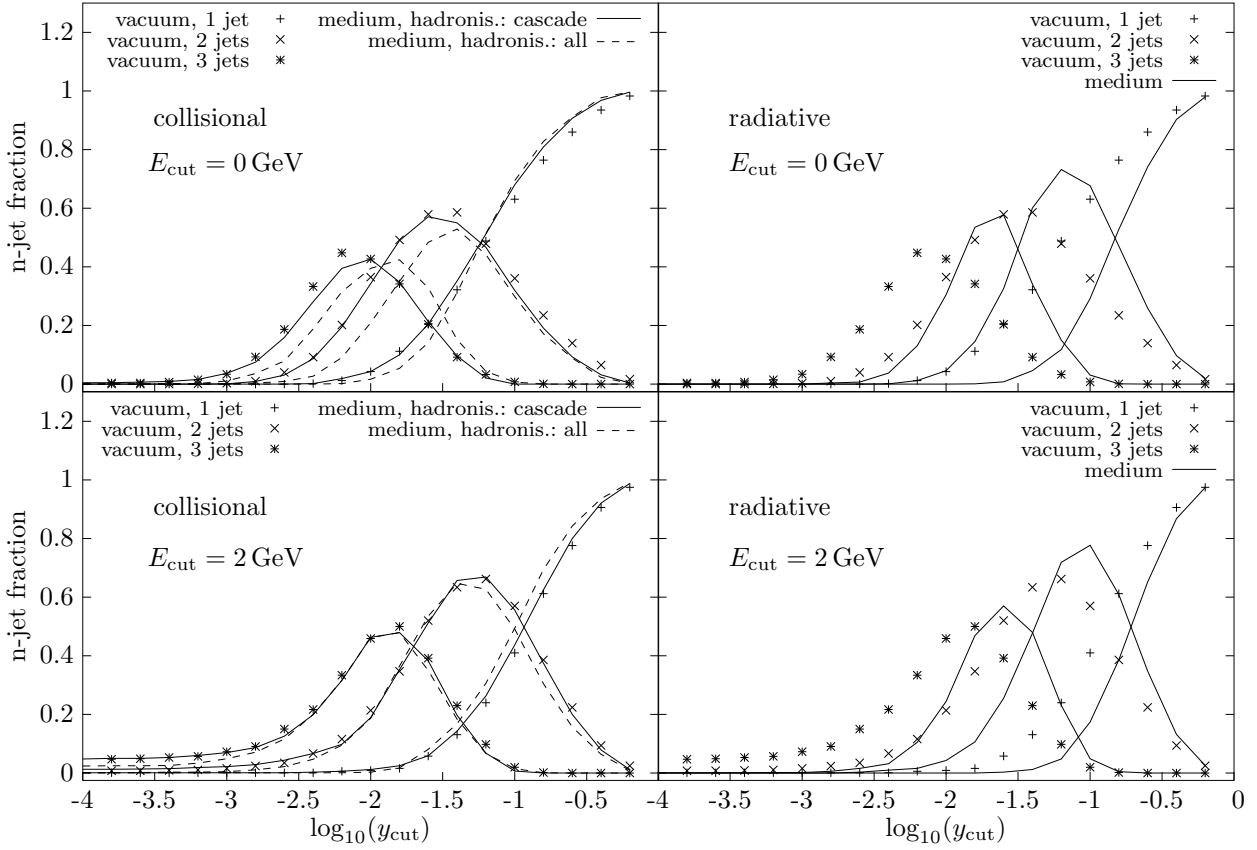
\begin{figure*}[t]
\centering
\input{8-njetmed-lqcd300.pstex_t}
\caption{$n$-jet fractions for a single \unit[100]{GeV} quark jet
after hadronisation
in vacuum and with medium effects. Left hand side: Collisional energy loss for
$T= \unit[500]{MeV}$
and $L= \unit[5]{fm}$. Recoil partons are either hadronised together with the
cascade
('all') or they are not included in the hadronisation ('cascade'). 
Right hand side: Medium-induced radiation for $f_{\rm med} =3$ and $L =
\unit[5]{fm}$. In the
top row, all hadrons are included, while in the bottom row, only hadrons with
energy
above $E_{\rm cut} = \unit[2]{GeV}$ are included.}
\label{fig9}
\end{figure*}
%

The medium-modification of the $n$-jet fraction may provide another tool for
disentangling
elastic from inelastic jet quenching mechanisms.
The potential impact of elastic $2\to 2$ processes on the
$n$-jet fraction is limited to two numerically rather minor effects:
First, elastic collisions are a source of angular broadening, which 
can increase the distance $y_{ij}$ between partons and this implies that
additional
jets can become visible on a coarser scale $y_{\rm cut}$. However, the higher
the
energy of the partonic projectile, the smaller is this broadening effect. For
this reason,
one expects the dependence of the $n$-jet fraction on collisional energy loss to
be
relatively small and mainly visible for relatively small values of $y_{\rm cut}$
which are
sensitive to small energies. Second, collisional mechanisms may kick target components towards higher
transverse momenta.
In this way, hard recoils may be counted as jets, thereby increasing the $n$-jet
fraction. However, from the momentum
space
distribution of recoil partners studied in figures ~\ref{fig6} and
~\ref{fig7},
we know that this effect will remain limited to very fine jet resolution
scales. This is confirmed by the results shown in
figure~\ref{fig9}, which shows
that even after an in-medium path length of $L=\unit[5]{fm}$ in a medium of
temperature
$T=\unit[500]{MeV}$, the $n$-jet fraction remains unaffected above ${\rm
log}_{10}(y_{\rm cut}) = -1.5$
and the mild medium-induced deviations seen at higher jet resolution scale can
be
attributed largely to soft hadrons with energy $E_\text{h} < E_{\rm cut} =
\unit[2]{GeV}$.

In contrast to elastic interactions, medium-induced radiation results in a 
distribution of sub-leading jet fragments up to characteristically larger
energies.
This is clearly seen in the medium-modified $n$-jet fractions of
figure~\ref{fig9}, which
shows a marked increase in the average number of jets even for a very coarse
resolution scale ${\rm log}_{10}(y_{\rm cut}) > -1$. After a background cut at
$E_{\rm cut} = \unit[2]{GeV}$, this medium-modification remains clearly
visible on the
logarithmic scale ${\rm log}_{10}(y_{\rm cut})$. 

We have checked that the $n$-jet fraction shows a very weak dependence on the
total jet energy in the range $75< E_{\rm jet} < \unit[125]{GeV}$.
However, if this jet energy is not measured
exactly, $E_{\rm jet\, ,true} = f_{\rm corr}\, E_{\rm jet\, ,meas}$, then the
$n$-jet fraction
is shifted by a constant value $-{\rm log}_{10} (f_{\rm corr}^2)$ in ${\rm
log}_{10}(y_{\rm cut}) $.
For instance, a deviation of the jet energy reconstructed from a calorimetric jet measurement from the true jet energy by \unit[30]{\%} amounts
to a shift of ${\rm log}_{10}(1.3^2) \approx 0.23$. This is comparable to the
size of the
medium modifications on the right hand side of figure~\ref{fig9}. So, our
studies
indicate that the $n$-jet fraction may provide a powerful tool for disentangling
radiative from
collisional energy loss mechanisms, but a proper assessment of various
experimental
uncertainties will be clearly needed to explore this tool. An analogous
statement applies
to the medium-modifications of jet shapes studied in figure~\ref{fig8}.

%
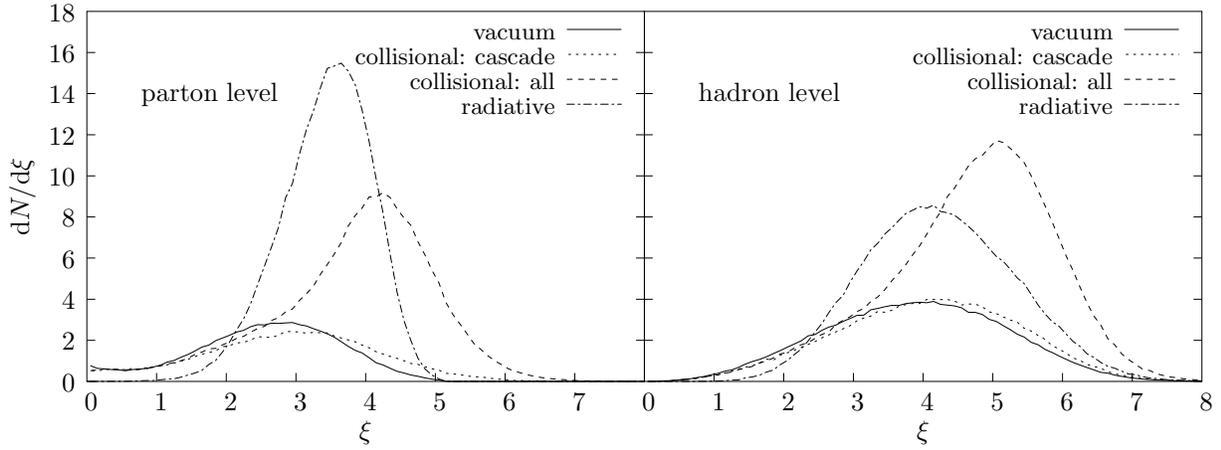
\begin{figure*}[t]
\centering
\input{7-ximed-lqcd300.pstex_t}
\caption{The single inclusive distribution $\d N/\d\xi$ for a single
medium-modified quark jet ($E_q = \unit[100]{GeV}$)
before and after hadronisation ($Q_0 = \unit[1]{GeV}$). On the parton level
(left hand side), all partons are shown, but on the hadron level (right hand side), 
only charged hadrons are included. Collisional energy loss is calculated 
for $T = \unit[500]{MeV}$ and $L = \unit[5]{fm}$, 
with recoil partons either hadronised together with the cascade
('all') or not included in the hadronisation ('cascade').
Medium induced radiation is calculated for $f_{\rm med} = 3$ and $L = \unit[5]{fm}$.}
\label{fig10}
\end{figure*}
%

In figure~\ref{fig10},  we plot the medium modification of the single inclusive
distribution
$\d N/\d\xi$ in the presence of collisional and radiative medium effects.
Similar to the case
for the vacuum shower in figure~\ref{fig3}, hadronisation is seen to lead to a
significant
softening of the distribution. If elastic interactions with the medium are
included but only the cascade is hadronised, then
the total jet multiplicity of projectile partons or hadrons 
depends only very weakly on the
medium, since $2\to 2$ processes do not increase the
parton multiplicity. However, the total jet multiplicity
may increase significantly if recoil partons are counted towards
the jet. This is illustrated in figure~\ref{fig10} which also shows the 
resulting $\d N/ \d\xi$ distribution after addition of all recoil partons that go in the same hemisphere as the original parton. Thus the jet-modification of the medium and the
medium-modification of the jet
are two complementary aspects of the same dynamic phenomenon.

As stated above, the present study of radiative parton energy loss mechanisms is
based on a simplified model which does not account for the recoil distribution of target
partons. However,
figure~\ref{fig10} indicates that radiative mechanisms
result in a larger
increase of intra-jet multiplicities. 

\section{Conclusion and Outlook}
Monte Carlo models for hadronic and nuclear collisions are at the interface
between
the theory of QCD and experiment. Depending on the class of measurements and the
status of theory, they provide a bridge between QCD and data which invokes a
varying
degree of model assumptions. In this paper, we have started to discuss the Monte
Carlo model \textsc{Jewel}, developed for a class of measurements sensitive to
jet quenching phenomena in ultra-relativistic nuclear collisions. In the absence
of
medium-effects, we have validated \textsc{Jewel} in section~\ref{sec2} against
a set of
benchmark jet measurements. In section~\ref{sec3}, we have then introduced
medium-modifications with a special focus on the dynamic description of 
elastic interactions between a fragmenting projectile and the medium. 
The main results, presented in section~\ref{sec4}, show that collisional energy loss is important and demonstrate the basic physical jet quenching effects for given temperature and path length in the plasma. Moreover, we find
that quantities like the $n$-jet fraction and jet event shapes may allow us to 
distinguish between elastic and inelastic mechanisms underlying medium-induced
parton energy loss. As discussed in the introduction, the microscopic mechanism(s) underlying jet 
quenching are not firmly established and their theoretical description is incomplete.
In view of the complexity of this problem in heavy ion collisions, the Monte Carlo simulation method is particularly suitable for a detailed treatment. 
Amongst the many open problems in the theory and phenomenology of jet quenching, we identify in particular the following points which may be accessible by further developments of \textsc{Jewel}:
\begin{enumerate}
         \item {\it Heavy flavour, multi-hadron correlations and other
measurements.}\\
         \textsc{Jewel} can be extended easily to include the in-medium
propagation of heavy quarks, which has attracted significant interest
recently~\cite{Djordjevic:2006tw,Adil:2006ei,Zakharov:2007pj,Peigne:2008nd,Zapp:2007zs}.
         Simulations of multi-hadron correlations can be extracted from the
present version
         of \textsc{Jewel}, but are likely to depend strongly on the
hadronisation model,
         as discussed in the context of figure~\ref{fig3}.
	\item {\it Realistic description of collision geometry and expansion.}\\
	On a schematic level, \textsc{Jewel} can be interfaced with any
simulation of the bulk
	evolution of heavy ion collisions, by supplementing \textsc{Jewel} with
information about
	a realistic distribution of partonic trajectories inside the medium, and
with information about a realistic
	time-dependent density of scattering centres along these trajectories.
In particular, this includes full 2-di\-men\-sio\-nal ideal~\cite{Kolb:2001qz}
or viscous~\cite{Song:2007ux,Romatschke:2007mq} and full 
3-di\-men\-sio\-nal ideal~\cite{Renk:2006sx} hydrodynamic
simulations of the heavy ion collision, as well as other models for an 
expanding plasma~\cite{Zapp:2005kt}.
	\item {\it Radiative parton energy loss.}\\
	The version of \textsc{Jewel} presented here mimics radiative energy loss
by enhancing
	the perturbative splitting functions. We regard the substitution of this
 {\it ad hoc} prescription by
	realistic $2\to 3$ inelastic parton scattering cross sections as a
feasible and interesting
	future extension of \textsc{Jewel}. It would provide for a radiative
energy loss mechanism
	which includes recoil effects, and it could define the incoherent limit of
any radiative
	mechanism. 
	\item {\it The separation between weakly coupled and strongly coupled
regimes and
	the problem of hadronisation.}\\
	On theoretical grounds, there is ample motivation to use weak
	coupling techniques for describing the splitting
	of highly virtual partons, and their
sufficiently hard interactions
	with a medium. However, for soft interactions with the medium, one has to invoke a
	non-perturbative mechanism for regularising the infra-red singularities,
resulting in
	uncertainties discussed in section~\ref{sec3}. Additional complications
arise
	if one considers the propagation of soft projectile fragments whose
momentum is
	close to that of components of the heat bath. Propagation of these
fragments within
	a perturbative parton cascade is expected to be unreliable since 
	the medium is likely to be strongly coupled and the dynamics of
the medium
	 cannot be accounted for by partonic $2\to 2$ and $2 \to 3$ processes
only. Another open question is how to hadronise these soft jet components
within
	 a soft high multiplicity environment, where novel hadronisation
mechanisms
	 such as recombination may become relevant. These processes may become a 
	 test laboratory for understanding how a well-defined partonic projectile
interacts
	 and to what extent it thermalises within a finite size medium.
	 From a pragmatic point of view, a parton shower may contribute to this
	 issue by identifying the momentum scales at which a perturbative
description
	 breaks down.
\end{enumerate}

There are many other open issues for improving our understanding of jet
quenching in
an interplay between theory and experiment. One of the most central ones may be
how
to best characterise a jet within a high multiplicity environment such that
unambiguous
information about its medium modifications can be obtained. Standard
calorimetric jet
measurements are very difficult to apply to heavy ion collisions, though novel
jet algorithms may be better suited~\cite{Cacciari:2005hq}. On the other
hand, the
suppression of single inclusive spectra
provides unambiguous evidence that there are strong medium modifications. There
is no fundamental reason why a similar unambiguous characterisation should
not be
possible for other aspects of the entire jet fragmentation pattern. From a
pragmatic
point of view, the question is then how to identify classes of jet measurements,
which are
sensitive to medium effects but which remain sufficiently insensitive to
operational uncertainties
in the jet definition. \textsc{Jewel} may contribute to this central issue on
various levels.
By superimposing simulations of \textsc{Jewel} on top of the simulated background of  heavy ion collisions (or developing \textsc{Jewel} to include this)
one can test the sensitivity of different jet observables. On the other hand, by simulating the redistribution
of 'background' multiplicity due to the propagation of a jet, \textsc{Jewel}
provides a mean to
go beyond the simplifying assumption that the medium-modified jet is
uncorrelated to
the underlying background. These features of  \textsc{Jewel} may contribute 
to establish to what extent
an operational procedure of characterising jet medium modifications
is suited to draw model-independent conclusions. 

\begin{acknowledgement}
We are greatly indebted to Peter Skands and Torbj\"orn Sj\"ostrand for numerous
discussions. We also acknowledge helpful discussions with G\"unther Dissertori,
Frank Krauss, Andre Peshier, Hans-J\"urgen Pirner and Peter Richardson. Korinna Zapp acknowledges 
support via a Marie Curie Early
Stage Research Training Fellowship of the European Community's Sixth Framework
Programme under contract number (MEST-CT-2005-\\020238-EUROTHEPHY). 
We also acknowledge support by the
German BMBF and the Swedish Research Council.
\end{acknowledgement}


\end{document}

%% file: 1-thrustvac-lqcd300.pstex_t
\begin{picture}(0,0)%
\includegraphics{1-thrustvac-lqcd300.pstex}%
\end{picture}%
\setlength{\unitlength}{3108sp}%
\begingroup\makeatletter\ifx\SetFigFontNFSS\undefined%
\gdef\SetFigFontNFSS#1#2#3#4#5{%
  \reset@font\fontsize{#1}{#2pt}%
  \fontfamily{#3}\fontseries{#4}\fontshape{#5}%
  \selectfont}%
\fi\endgroup%
\begin{picture}(5183,3506)(948,-3739)
\put(5329,-1246){\makebox(0,0)[rb]{\smash{{\SetFigFontNFSS{9}{10.8}{\familydefault}{\mddefault}{\updefault}hadron level, $Q_0=\unit[1]{GeV}$}}}}
\put(5329,-1051){\makebox(0,0)[rb]{\smash{{\SetFigFontNFSS{9}{10.8}{\familydefault}{\mddefault}{\updefault}parton level, $Q_0=\unit[1]{GeV}$}}}}
\put(5329,-871){\makebox(0,0)[rb]{\smash{{\SetFigFontNFSS{9}{10.8}{\familydefault}{\mddefault}{\updefault}$T_\text{min}$, \textsc{Aleph} data}}}}
\put(5329,-665){\makebox(0,0)[rb]{\smash{{\SetFigFontNFSS{9}{10.8}{\familydefault}{\mddefault}{\updefault}$T_\text{maj}$, \textsc{Aleph} data}}}}
\put(5329,-481){\makebox(0,0)[rb]{\smash{{\SetFigFontNFSS{9}{10.8}{\familydefault}{\mddefault}{\updefault}$T$, \textsc{Aleph} data}}}}
\put(1592,-3276){\makebox(0,0)[rb]{\smash{{\SetFigFontNFSS{10}{12.0}{\familydefault}{\mddefault}{\updefault} 0.1}}}}
\put(1592,-2381){\makebox(0,0)[rb]{\smash{{\SetFigFontNFSS{10}{12.0}{\familydefault}{\mddefault}{\updefault} 1}}}}
\put(1592,-1486){\makebox(0,0)[rb]{\smash{{\SetFigFontNFSS{10}{12.0}{\familydefault}{\mddefault}{\updefault} 10}}}}
\put(1592,-591){\makebox(0,0)[rb]{\smash{{\SetFigFontNFSS{10}{12.0}{\familydefault}{\mddefault}{\updefault} 100}}}}
\put(1684,-3430){\makebox(0,0)[b]{\smash{{\SetFigFontNFSS{10}{12.0}{\familydefault}{\mddefault}{\updefault} 0}}}}
\put(2555,-3430){\makebox(0,0)[b]{\smash{{\SetFigFontNFSS{10}{12.0}{\familydefault}{\mddefault}{\updefault} 0.2}}}}
\put(3426,-3430){\makebox(0,0)[b]{\smash{{\SetFigFontNFSS{10}{12.0}{\familydefault}{\mddefault}{\updefault} 0.4}}}}
\put(4298,-3430){\makebox(0,0)[b]{\smash{{\SetFigFontNFSS{10}{12.0}{\familydefault}{\mddefault}{\updefault} 0.6}}}}
\put(5169,-3430){\makebox(0,0)[b]{\smash{{\SetFigFontNFSS{10}{12.0}{\familydefault}{\mddefault}{\updefault} 0.8}}}}
\put(6040,-3430){\makebox(0,0)[b]{\smash{{\SetFigFontNFSS{10}{12.0}{\familydefault}{\mddefault}{\updefault} 1}}}}
\put(1131,-1722){\rotatebox{90.0}{\makebox(0,0)[b]{\smash{{\SetFigFontNFSS{10}{12.0}{\familydefault}{\mddefault}{\updefault}$\d N/\d T_\text{r}$}}}}}
\put(3862,-3661){\makebox(0,0)[b]{\smash{{\SetFigFontNFSS{10}{12.0}{\familydefault}{\mddefault}{\updefault}$T_\text{r}$}}}}
\end{picture}%

%% file: 2-njetvac-lqcd300.pstex_t
\begin{picture}(0,0)%
\includegraphics{2-njetvac-lqcd300.pstex}%
\end{picture}%
\setlength{\unitlength}{3108sp}%
\begingroup\makeatletter\ifx\SetFigFontNFSS\undefined%
\gdef\SetFigFontNFSS#1#2#3#4#5{%
  \reset@font\fontsize{#1}{#2pt}%
  \fontfamily{#3}\fontseries{#4}\fontshape{#5}%
  \selectfont}%
\fi\endgroup%
\begin{picture}(9469,3582)(948,-3739)
\put(9784,-876){\makebox(0,0)[rb]{\smash{{\SetFigFontNFSS{9}{10.8}{\familydefault}{\mddefault}{\updefault}hadron level}}}}
\put(9784,-677){\makebox(0,0)[rb]{\smash{{\SetFigFontNFSS{9}{10.8}{\familydefault}{\mddefault}{\updefault}$\ge 6$ jets \textsc{Aleph} data}}}}
\put(5419,-478){\makebox(0,0)[rb]{\smash{{\SetFigFontNFSS{9}{10.8}{\familydefault}{\mddefault}{\updefault}5 jets, parton level}}}}
\put(5419,-677){\makebox(0,0)[rb]{\smash{{\SetFigFontNFSS{9}{10.8}{\familydefault}{\mddefault}{\updefault}$\ge 6$ jets, parton level}}}}
\put(5419,-876){\makebox(0,0)[rb]{\smash{{\SetFigFontNFSS{9}{10.8}{\familydefault}{\mddefault}{\updefault}hadron level}}}}
\put(7669,-864){\makebox(0,0)[rb]{\smash{{\SetFigFontNFSS{9}{10.8}{\familydefault}{\mddefault}{\updefault}4 jets \textsc{Aleph} data}}}}
\put(7669,-665){\makebox(0,0)[rb]{\smash{{\SetFigFontNFSS{9}{10.8}{\familydefault}{\mddefault}{\updefault}3 jets \textsc{Aleph} data}}}}
\put(7669,-466){\makebox(0,0)[rb]{\smash{{\SetFigFontNFSS{9}{10.8}{\familydefault}{\mddefault}{\updefault}2 jets \textsc{Aleph} data}}}}
\put(9784,-478){\makebox(0,0)[rb]{\smash{{\SetFigFontNFSS{9}{10.8}{\familydefault}{\mddefault}{\updefault}5 jets \textsc{Aleph} data}}}}
\put(6122,-3430){\makebox(0,0)[b]{\smash{{\SetFigFontNFSS{10}{12.0}{\familydefault}{\mddefault}{\updefault}-6}}}}
\put(6848,-3430){\makebox(0,0)[b]{\smash{{\SetFigFontNFSS{10}{12.0}{\familydefault}{\mddefault}{\updefault}-5}}}}
\put(7574,-3430){\makebox(0,0)[b]{\smash{{\SetFigFontNFSS{10}{12.0}{\familydefault}{\mddefault}{\updefault}-4}}}}
\put(8300,-3430){\makebox(0,0)[b]{\smash{{\SetFigFontNFSS{10}{12.0}{\familydefault}{\mddefault}{\updefault}-3}}}}
\put(9026,-3430){\makebox(0,0)[b]{\smash{{\SetFigFontNFSS{10}{12.0}{\familydefault}{\mddefault}{\updefault}-2}}}}
\put(9752,-3430){\makebox(0,0)[b]{\smash{{\SetFigFontNFSS{10}{12.0}{\familydefault}{\mddefault}{\updefault}-1}}}}
\put(8227,-3661){\makebox(0,0)[b]{\smash{{\SetFigFontNFSS{10}{12.0}{\familydefault}{\mddefault}{\updefault}$\log_{10}(y_\text{cut})$}}}}
\put(3346,-864){\makebox(0,0)[rb]{\smash{{\SetFigFontNFSS{9}{10.8}{\familydefault}{\mddefault}{\updefault}4 jets, parton level}}}}
\put(3346,-665){\makebox(0,0)[rb]{\smash{{\SetFigFontNFSS{9}{10.8}{\familydefault}{\mddefault}{\updefault}3 jets, parton level}}}}
\put(3346,-466){\makebox(0,0)[rb]{\smash{{\SetFigFontNFSS{9}{10.8}{\familydefault}{\mddefault}{\updefault}2 jets, parton level}}}}
\put(1592,-3224){\makebox(0,0)[rb]{\smash{{\SetFigFontNFSS{10}{12.0}{\familydefault}{\mddefault}{\updefault} 0}}}}
\put(1592,-2810){\makebox(0,0)[rb]{\smash{{\SetFigFontNFSS{10}{12.0}{\familydefault}{\mddefault}{\updefault} 0.2}}}}
\put(1592,-2395){\makebox(0,0)[rb]{\smash{{\SetFigFontNFSS{10}{12.0}{\familydefault}{\mddefault}{\updefault} 0.4}}}}
\put(1592,-1980){\makebox(0,0)[rb]{\smash{{\SetFigFontNFSS{10}{12.0}{\familydefault}{\mddefault}{\updefault} 0.6}}}}
\put(1592,-1566){\makebox(0,0)[rb]{\smash{{\SetFigFontNFSS{10}{12.0}{\familydefault}{\mddefault}{\updefault} 0.8}}}}
\put(1592,-1151){\makebox(0,0)[rb]{\smash{{\SetFigFontNFSS{10}{12.0}{\familydefault}{\mddefault}{\updefault} 1}}}}
\put(1592,-737){\makebox(0,0)[rb]{\smash{{\SetFigFontNFSS{10}{12.0}{\familydefault}{\mddefault}{\updefault} 1.2}}}}
\put(1592,-322){\makebox(0,0)[rb]{\smash{{\SetFigFontNFSS{10}{12.0}{\familydefault}{\mddefault}{\updefault} 1.4}}}}
\put(1757,-3430){\makebox(0,0)[b]{\smash{{\SetFigFontNFSS{10}{12.0}{\familydefault}{\mddefault}{\updefault}-6}}}}
\put(2483,-3430){\makebox(0,0)[b]{\smash{{\SetFigFontNFSS{10}{12.0}{\familydefault}{\mddefault}{\updefault}-5}}}}
\put(3209,-3430){\makebox(0,0)[b]{\smash{{\SetFigFontNFSS{10}{12.0}{\familydefault}{\mddefault}{\updefault}-4}}}}
\put(3935,-3430){\makebox(0,0)[b]{\smash{{\SetFigFontNFSS{10}{12.0}{\familydefault}{\mddefault}{\updefault}-3}}}}
\put(4661,-3430){\makebox(0,0)[b]{\smash{{\SetFigFontNFSS{10}{12.0}{\familydefault}{\mddefault}{\updefault}-2}}}}
\put(5387,-3430){\makebox(0,0)[b]{\smash{{\SetFigFontNFSS{10}{12.0}{\familydefault}{\mddefault}{\updefault}-1}}}}
\put(1131,-1722){\rotatebox{90.0}{\makebox(0,0)[b]{\smash{{\SetFigFontNFSS{10}{12.0}{\familydefault}{\mddefault}{\updefault}$n$-jet fraction}}}}}
\put(3862,-3661){\makebox(0,0)[b]{\smash{{\SetFigFontNFSS{10}{12.0}{\familydefault}{\mddefault}{\updefault}$\log_{10}(y_\text{cut})$}}}}
\end{picture}%

%% file: 3-xivac-lqcd300.pstex_t
\begin{picture}(0,0)%
\includegraphics{3-xivac-lqcd300.pstex}%
\end{picture}%
\setlength{\unitlength}{3108sp}%
\begingroup\makeatletter\ifx\SetFigFontNFSS\undefined%
\gdef\SetFigFontNFSS#1#2#3#4#5{%
  \reset@font\fontsize{#1}{#2pt}%
  \fontfamily{#3}\fontseries{#4}\fontshape{#5}%
  \selectfont}%
\fi\endgroup%
\begin{picture}(5092,3497)(960,-3730)
\put(5329,-864){\makebox(0,0)[rb]{\smash{{\SetFigFontNFSS{9}{10.8}{\familydefault}{\mddefault}{\updefault}hadron level, $Q_0=\unit[1]{GeV}$}}}}
\put(5329,-665){\makebox(0,0)[rb]{\smash{{\SetFigFontNFSS{9}{10.8}{\familydefault}{\mddefault}{\updefault}parton level, $Q_0=\unit[1]{GeV}$}}}}
\put(5329,-466){\makebox(0,0)[rb]{\smash{{\SetFigFontNFSS{9}{10.8}{\familydefault}{\mddefault}{\updefault}\textsc{Aleph} data}}}}
\put(1500,-3276){\makebox(0,0)[rb]{\smash{{\SetFigFontNFSS{10}{12.0}{\familydefault}{\mddefault}{\updefault} 0}}}}
\put(1500,-2822){\makebox(0,0)[rb]{\smash{{\SetFigFontNFSS{10}{12.0}{\familydefault}{\mddefault}{\updefault} 2}}}}
\put(1500,-2367){\makebox(0,0)[rb]{\smash{{\SetFigFontNFSS{10}{12.0}{\familydefault}{\mddefault}{\updefault} 4}}}}
\put(1500,-1913){\makebox(0,0)[rb]{\smash{{\SetFigFontNFSS{10}{12.0}{\familydefault}{\mddefault}{\updefault} 6}}}}
\put(1500,-1458){\makebox(0,0)[rb]{\smash{{\SetFigFontNFSS{10}{12.0}{\familydefault}{\mddefault}{\updefault} 8}}}}
\put(1500,-1004){\makebox(0,0)[rb]{\smash{{\SetFigFontNFSS{10}{12.0}{\familydefault}{\mddefault}{\updefault} 10}}}}
\put(1500,-549){\makebox(0,0)[rb]{\smash{{\SetFigFontNFSS{10}{12.0}{\familydefault}{\mddefault}{\updefault} 12}}}}
\put(1592,-3430){\makebox(0,0)[b]{\smash{{\SetFigFontNFSS{10}{12.0}{\familydefault}{\mddefault}{\updefault} 0}}}}
\put(2185,-3430){\makebox(0,0)[b]{\smash{{\SetFigFontNFSS{10}{12.0}{\familydefault}{\mddefault}{\updefault} 1}}}}
\put(2778,-3430){\makebox(0,0)[b]{\smash{{\SetFigFontNFSS{10}{12.0}{\familydefault}{\mddefault}{\updefault} 2}}}}
\put(3371,-3430){\makebox(0,0)[b]{\smash{{\SetFigFontNFSS{10}{12.0}{\familydefault}{\mddefault}{\updefault} 3}}}}
\put(3964,-3430){\makebox(0,0)[b]{\smash{{\SetFigFontNFSS{10}{12.0}{\familydefault}{\mddefault}{\updefault} 4}}}}
\put(4557,-3430){\makebox(0,0)[b]{\smash{{\SetFigFontNFSS{10}{12.0}{\familydefault}{\mddefault}{\updefault} 5}}}}
\put(5150,-3430){\makebox(0,0)[b]{\smash{{\SetFigFontNFSS{10}{12.0}{\familydefault}{\mddefault}{\updefault} 6}}}}
\put(5743,-3430){\makebox(0,0)[b]{\smash{{\SetFigFontNFSS{10}{12.0}{\familydefault}{\mddefault}{\updefault} 7}}}}
\put(1131,-1722){\rotatebox{90.0}{\makebox(0,0)[b]{\smash{{\SetFigFontNFSS{10}{12.0}{\familydefault}{\mddefault}{\updefault}$\d N/\d\xi$}}}}}
\put(3816,-3661){\makebox(0,0)[b]{\smash{{\SetFigFontNFSS{10}{12.0}{\familydefault}{\mddefault}{\updefault}$\xi$}}}}
\end{picture}%

%% file: 4-dedx-lqcd300.pstex_t
\begin{picture}(0,0)%
\includegraphics{4-dedx-lqcd300.pstex}%
\end{picture}%
\setlength{\unitlength}{3108sp}%
\begingroup\makeatletter\ifx\SetFigFontNFSS\undefined%
\gdef\SetFigFontNFSS#1#2#3#4#5{%
  \reset@font\fontsize{#1}{#2pt}%
  \fontfamily{#3}\fontseries{#4}\fontshape{#5}%
  \selectfont}%
\fi\endgroup%
\begin{picture}(5183,3582)(948,-3739)
\put(2143,-466){\makebox(0,0)[lb]{\smash{{\SetFigFontNFSS{9}{10.8}{\familydefault}{\mddefault}{\updefault}cross section I, $E=\unit[10]{GeV}$}}}}
\put(2143,-665){\makebox(0,0)[lb]{\smash{{\SetFigFontNFSS{9}{10.8}{\familydefault}{\mddefault}{\updefault}cross section II, $E=\unit[10]{GeV}$}}}}
\put(2143,-883){\makebox(0,0)[lb]{\smash{{\SetFigFontNFSS{9}{10.8}{\familydefault}{\mddefault}{\updefault}cross section I, $E=\unit[100]{GeV}$}}}}
\put(2143,-1082){\makebox(0,0)[lb]{\smash{{\SetFigFontNFSS{9}{10.8}{\familydefault}{\mddefault}{\updefault}cross section II, $E=\unit[100]{GeV}$}}}}
\put(1408,-3276){\makebox(0,0)[rb]{\smash{{\SetFigFontNFSS{10}{12.0}{\familydefault}{\mddefault}{\updefault} 0}}}}
\put(1408,-2685){\makebox(0,0)[rb]{\smash{{\SetFigFontNFSS{10}{12.0}{\familydefault}{\mddefault}{\updefault} 1}}}}
\put(1408,-2094){\makebox(0,0)[rb]{\smash{{\SetFigFontNFSS{10}{12.0}{\familydefault}{\mddefault}{\updefault} 2}}}}
\put(1408,-1504){\makebox(0,0)[rb]{\smash{{\SetFigFontNFSS{10}{12.0}{\familydefault}{\mddefault}{\updefault} 3}}}}
\put(1408,-913){\makebox(0,0)[rb]{\smash{{\SetFigFontNFSS{10}{12.0}{\familydefault}{\mddefault}{\updefault} 4}}}}
\put(1408,-322){\makebox(0,0)[rb]{\smash{{\SetFigFontNFSS{10}{12.0}{\familydefault}{\mddefault}{\updefault} 5}}}}
\put(1500,-3430){\makebox(0,0)[b]{\smash{{\SetFigFontNFSS{10}{12.0}{\familydefault}{\mddefault}{\updefault} 0.2}}}}
\put(2068,-3430){\makebox(0,0)[b]{\smash{{\SetFigFontNFSS{10}{12.0}{\familydefault}{\mddefault}{\updefault} 0.3}}}}
\put(2635,-3430){\makebox(0,0)[b]{\smash{{\SetFigFontNFSS{10}{12.0}{\familydefault}{\mddefault}{\updefault} 0.4}}}}
\put(3202,-3430){\makebox(0,0)[b]{\smash{{\SetFigFontNFSS{10}{12.0}{\familydefault}{\mddefault}{\updefault} 0.5}}}}
\put(3770,-3430){\makebox(0,0)[b]{\smash{{\SetFigFontNFSS{10}{12.0}{\familydefault}{\mddefault}{\updefault} 0.6}}}}
\put(4337,-3430){\makebox(0,0)[b]{\smash{{\SetFigFontNFSS{10}{12.0}{\familydefault}{\mddefault}{\updefault} 0.7}}}}
\put(4905,-3430){\makebox(0,0)[b]{\smash{{\SetFigFontNFSS{10}{12.0}{\familydefault}{\mddefault}{\updefault} 0.8}}}}
\put(5472,-3430){\makebox(0,0)[b]{\smash{{\SetFigFontNFSS{10}{12.0}{\familydefault}{\mddefault}{\updefault} 0.9}}}}
\put(6040,-3430){\makebox(0,0)[b]{\smash{{\SetFigFontNFSS{10}{12.0}{\familydefault}{\mddefault}{\updefault} 1}}}}
\put(1131,-1722){\rotatebox{90.0}{\makebox(0,0)[b]{\smash{{\SetFigFontNFSS{10}{12.0}{\familydefault}{\mddefault}{\updefault}$-\d E/\d x\ [\unit{GeV/fm}]$}}}}}
\put(3770,-3661){\makebox(0,0)[b]{\smash{{\SetFigFontNFSS{10}{12.0}{\familydefault}{\mddefault}{\updefault}$T\ [\unit{GeV}]$}}}}
\end{picture}%

%% file: 5-deltae-lqcd300.pstex_t
\begin{picture}(0,0)%
\includegraphics{5-deltae-lqcd300.pstex}%
\end{picture}%
\setlength{\unitlength}{3108sp}%
\begingroup\makeatletter\ifx\SetFigFontNFSS\undefined%
\gdef\SetFigFontNFSS#1#2#3#4#5{%
  \reset@font\fontsize{#1}{#2pt}%
  \fontfamily{#3}\fontseries{#4}\fontshape{#5}%
  \selectfont}%
\fi\endgroup%
\begin{picture}(9416,6459)(948,-6692)
\put(8461,-1546){\makebox(0,0)[lb]{\smash{{\SetFigFontNFSS{10}{12.0}{\familydefault}{\mddefault}{\updefault}$E=\unit[100]{GeV}$}}}}
\put(4366,-4948){\makebox(0,0)[lb]{\smash{{\SetFigFontNFSS{10}{12.0}{\familydefault}{\mddefault}{\updefault}$E=\unit[10]{GeV}$}}}}
\put(8455,-5174){\makebox(0,0)[lb]{\smash{{\SetFigFontNFSS{10}{12.0}{\familydefault}{\mddefault}{\updefault}$E=\unit[100]{GeV}$}}}}
\put(4276,-1546){\makebox(0,0)[lb]{\smash{{\SetFigFontNFSS{10}{12.0}{\familydefault}{\mddefault}{\updefault}$E=\unit[10]{GeV}$}}}}
\put(5329,-1063){\makebox(0,0)[rb]{\smash{{\SetFigFontNFSS{9}{10.8}{\familydefault}{\mddefault}{\updefault}cross section II, $T=\unit[500]{MeV}$}}}}
\put(5329,-864){\makebox(0,0)[rb]{\smash{{\SetFigFontNFSS{9}{10.8}{\familydefault}{\mddefault}{\updefault}cross section I, $T=\unit[500]{MeV}$}}}}
\put(5329,-665){\makebox(0,0)[rb]{\smash{{\SetFigFontNFSS{9}{10.8}{\familydefault}{\mddefault}{\updefault}cross section II, $T=\unit[200]{MeV}$}}}}
\put(5329,-466){\makebox(0,0)[rb]{\smash{{\SetFigFontNFSS{9}{10.8}{\familydefault}{\mddefault}{\updefault}cross section I, $T=\unit[200]{MeV}$}}}}
\put(9514,-1063){\makebox(0,0)[rb]{\smash{{\SetFigFontNFSS{9}{10.8}{\familydefault}{\mddefault}{\updefault}cross section II, $T=\unit[500]{MeV}$}}}}
\put(9514,-864){\makebox(0,0)[rb]{\smash{{\SetFigFontNFSS{9}{10.8}{\familydefault}{\mddefault}{\updefault}cross section I, $T=\unit[500]{MeV}$}}}}
\put(9514,-665){\makebox(0,0)[rb]{\smash{{\SetFigFontNFSS{9}{10.8}{\familydefault}{\mddefault}{\updefault}cross section II, $T=\unit[200]{MeV}$}}}}
\put(9514,-466){\makebox(0,0)[rb]{\smash{{\SetFigFontNFSS{9}{10.8}{\familydefault}{\mddefault}{\updefault}cross section I, $T=\unit[200]{MeV}$}}}}
\put(5329,-3816){\makebox(0,0)[rb]{\smash{{\SetFigFontNFSS{9}{10.8}{\familydefault}{\mddefault}{\updefault}cross section I, $T=\unit[500]{MeV}$, $L=\unit[5]{fm}$}}}}
\put(5329,-3617){\makebox(0,0)[rb]{\smash{{\SetFigFontNFSS{9}{10.8}{\familydefault}{\mddefault}{\updefault}cross section I, $T=\unit[500]{MeV}$, $L=\unit[3]{fm}$}}}}
\put(5329,-3418){\makebox(0,0)[rb]{\smash{{\SetFigFontNFSS{9}{10.8}{\familydefault}{\mddefault}{\updefault}cross section I, $T=\unit[500]{MeV}$, $L=\unit[1]{fm}$}}}}
\put(1776,-6228){\makebox(0,0)[rb]{\smash{{\SetFigFontNFSS{10}{12.0}{\familydefault}{\mddefault}{\updefault} 0}}}}
\put(1776,-5834){\makebox(0,0)[rb]{\smash{{\SetFigFontNFSS{10}{12.0}{\familydefault}{\mddefault}{\updefault} 0.2}}}}
\put(1776,-5440){\makebox(0,0)[rb]{\smash{{\SetFigFontNFSS{10}{12.0}{\familydefault}{\mddefault}{\updefault} 0.4}}}}
\put(1776,-5046){\makebox(0,0)[rb]{\smash{{\SetFigFontNFSS{10}{12.0}{\familydefault}{\mddefault}{\updefault} 0.6}}}}
\put(1776,-4653){\makebox(0,0)[rb]{\smash{{\SetFigFontNFSS{10}{12.0}{\familydefault}{\mddefault}{\updefault} 0.8}}}}
\put(1776,-4259){\makebox(0,0)[rb]{\smash{{\SetFigFontNFSS{10}{12.0}{\familydefault}{\mddefault}{\updefault} 1}}}}
\put(1776,-3865){\makebox(0,0)[rb]{\smash{{\SetFigFontNFSS{10}{12.0}{\familydefault}{\mddefault}{\updefault} 1.2}}}}
\put(1776,-3471){\makebox(0,0)[rb]{\smash{{\SetFigFontNFSS{10}{12.0}{\familydefault}{\mddefault}{\updefault} 1.4}}}}
\put(2080,-6382){\makebox(0,0)[b]{\smash{{\SetFigFontNFSS{10}{12.0}{\familydefault}{\mddefault}{\updefault} 0}}}}
\put(2872,-6382){\makebox(0,0)[b]{\smash{{\SetFigFontNFSS{10}{12.0}{\familydefault}{\mddefault}{\updefault} 2}}}}
\put(3664,-6382){\makebox(0,0)[b]{\smash{{\SetFigFontNFSS{10}{12.0}{\familydefault}{\mddefault}{\updefault} 4}}}}
\put(4456,-6382){\makebox(0,0)[b]{\smash{{\SetFigFontNFSS{10}{12.0}{\familydefault}{\mddefault}{\updefault} 6}}}}
\put(5248,-6382){\makebox(0,0)[b]{\smash{{\SetFigFontNFSS{10}{12.0}{\familydefault}{\mddefault}{\updefault} 8}}}}
\put(6040,-6382){\makebox(0,0)[b]{\smash{{\SetFigFontNFSS{10}{12.0}{\familydefault}{\mddefault}{\updefault} 10}}}}
\put(1131,-4674){\rotatebox{90.0}{\makebox(0,0)[b]{\smash{{\SetFigFontNFSS{10}{12.0}{\familydefault}{\mddefault}{\updefault}$P(\text{energy loss} \le \Delta E)$}}}}}
\put(3862,-6613){\makebox(0,0)[b]{\smash{{\SetFigFontNFSS{10}{12.0}{\familydefault}{\mddefault}{\updefault}$\Delta E\ [\unit{GeV}]$}}}}
\put(9508,-3618){\makebox(0,0)[rb]{\smash{{\SetFigFontNFSS{9}{10.8}{\familydefault}{\mddefault}{\updefault}cross section I, $T=\unit[500]{MeV}$, $L=\unit[3]{fm}$}}}}
\put(9508,-3419){\makebox(0,0)[rb]{\smash{{\SetFigFontNFSS{9}{10.8}{\familydefault}{\mddefault}{\updefault}cross section I, $T=\unit[500]{MeV}$, $L=\unit[1]{fm}$}}}}
\put(9508,-3817){\makebox(0,0)[rb]{\smash{{\SetFigFontNFSS{9}{10.8}{\familydefault}{\mddefault}{\updefault}cross section I, $T=\unit[500]{MeV}$, $L=\unit[5]{fm}$}}}}
\put(6259,-6383){\makebox(0,0)[b]{\smash{{\SetFigFontNFSS{10}{12.0}{\familydefault}{\mddefault}{\updefault} 0}}}}
\put(7051,-6383){\makebox(0,0)[b]{\smash{{\SetFigFontNFSS{10}{12.0}{\familydefault}{\mddefault}{\updefault} 2}}}}
\put(7843,-6383){\makebox(0,0)[b]{\smash{{\SetFigFontNFSS{10}{12.0}{\familydefault}{\mddefault}{\updefault} 4}}}}
\put(8635,-6383){\makebox(0,0)[b]{\smash{{\SetFigFontNFSS{10}{12.0}{\familydefault}{\mddefault}{\updefault} 6}}}}
\put(9427,-6383){\makebox(0,0)[b]{\smash{{\SetFigFontNFSS{10}{12.0}{\familydefault}{\mddefault}{\updefault} 8}}}}
\put(10219,-6383){\makebox(0,0)[b]{\smash{{\SetFigFontNFSS{10}{12.0}{\familydefault}{\mddefault}{\updefault} 10}}}}
\put(8041,-6614){\makebox(0,0)[b]{\smash{{\SetFigFontNFSS{10}{12.0}{\familydefault}{\mddefault}{\updefault}$\Delta E\ [\unit{GeV}]$}}}}
\put(1776,-3276){\makebox(0,0)[rb]{\smash{{\SetFigFontNFSS{10}{12.0}{\familydefault}{\mddefault}{\updefault} 0.0001}}}}
\put(1776,-2835){\makebox(0,0)[rb]{\smash{{\SetFigFontNFSS{10}{12.0}{\familydefault}{\mddefault}{\updefault} 0.001}}}}
\put(1776,-2394){\makebox(0,0)[rb]{\smash{{\SetFigFontNFSS{10}{12.0}{\familydefault}{\mddefault}{\updefault} 0.01}}}}
\put(1776,-1953){\makebox(0,0)[rb]{\smash{{\SetFigFontNFSS{10}{12.0}{\familydefault}{\mddefault}{\updefault} 0.1}}}}
\put(1776,-1512){\makebox(0,0)[rb]{\smash{{\SetFigFontNFSS{10}{12.0}{\familydefault}{\mddefault}{\updefault} 1}}}}
\put(1776,-1071){\makebox(0,0)[rb]{\smash{{\SetFigFontNFSS{10}{12.0}{\familydefault}{\mddefault}{\updefault} 10}}}}
\put(1776,-630){\makebox(0,0)[rb]{\smash{{\SetFigFontNFSS{10}{12.0}{\familydefault}{\mddefault}{\updefault} 100}}}}
\put(1131,-1722){\rotatebox{90.0}{\makebox(0,0)[b]{\smash{{\SetFigFontNFSS{10}{12.0}{\familydefault}{\mddefault}{\updefault}$\d N/\d(\Delta E)$}}}}}
\end{picture}%

%% file: 6-bckgrnd-lqcd300.pstex_t
\begin{picture}(0,0)%
\includegraphics{6-bckgrnd-lqcd300.pstex}%
\end{picture}%
\setlength{\unitlength}{3108sp}%
\begingroup\makeatletter\ifx\SetFigFontNFSS\undefined%
\gdef\SetFigFontNFSS#1#2#3#4#5{%
  \reset@font\fontsize{#1}{#2pt}%
  \fontfamily{#3}\fontseries{#4}\fontshape{#5}%
  \selectfont}%
\fi\endgroup%
\begin{picture}(10120,6527)(948,-6684)
\put(6616,-1546){\makebox(0,0)[lb]{\smash{{\SetFigFontNFSS{10}{12.0}{\familydefault}{\mddefault}{\updefault}without splitting}}}}
\put(4276,-4491){\makebox(0,0)[lb]{\smash{{\SetFigFontNFSS{10}{12.0}{\familydefault}{\mddefault}{\updefault}with splitting}}}}
\put(6614,-4491){\makebox(0,0)[lb]{\smash{{\SetFigFontNFSS{10}{12.0}{\familydefault}{\mddefault}{\updefault}with splitting}}}}
\put(4276,-1501){\makebox(0,0)[lb]{\smash{{\SetFigFontNFSS{10}{12.0}{\familydefault}{\mddefault}{\updefault}without splitting}}}}
\put(9694,-1063){\makebox(0,0)[rb]{\smash{{\SetFigFontNFSS{9}{10.8}{\familydefault}{\mddefault}{\updefault}$T=\unit[500]{MeV}$, recoil}}}}
\put(9694,-864){\makebox(0,0)[rb]{\smash{{\SetFigFontNFSS{9}{10.8}{\familydefault}{\mddefault}{\updefault}$T=\unit[500]{MeV}$, medium}}}}
\put(9694,-665){\makebox(0,0)[rb]{\smash{{\SetFigFontNFSS{9}{10.8}{\familydefault}{\mddefault}{\updefault}$T=\unit[200]{MeV}$, recoil}}}}
\put(9694,-466){\makebox(0,0)[rb]{\smash{{\SetFigFontNFSS{9}{10.8}{\familydefault}{\mddefault}{\updefault}$T=\unit[200]{MeV}$, medium}}}}
\put(10425,-3276){\makebox(0,0)[lb]{\smash{{\SetFigFontNFSS{10}{12.0}{\familydefault}{\mddefault}{\updefault} 0}}}}
\put(10425,-2907){\makebox(0,0)[lb]{\smash{{\SetFigFontNFSS{10}{12.0}{\familydefault}{\mddefault}{\updefault} 0.2}}}}
\put(10425,-2537){\makebox(0,0)[lb]{\smash{{\SetFigFontNFSS{10}{12.0}{\familydefault}{\mddefault}{\updefault} 0.4}}}}
\put(10425,-2168){\makebox(0,0)[lb]{\smash{{\SetFigFontNFSS{10}{12.0}{\familydefault}{\mddefault}{\updefault} 0.6}}}}
\put(10425,-1799){\makebox(0,0)[lb]{\smash{{\SetFigFontNFSS{10}{12.0}{\familydefault}{\mddefault}{\updefault} 0.8}}}}
\put(10425,-1430){\makebox(0,0)[lb]{\smash{{\SetFigFontNFSS{10}{12.0}{\familydefault}{\mddefault}{\updefault} 1}}}}
\put(10425,-1061){\makebox(0,0)[lb]{\smash{{\SetFigFontNFSS{10}{12.0}{\familydefault}{\mddefault}{\updefault} 1.2}}}}
\put(10425,-691){\makebox(0,0)[lb]{\smash{{\SetFigFontNFSS{10}{12.0}{\familydefault}{\mddefault}{\updefault} 1.4}}}}
\put(10425,-322){\makebox(0,0)[lb]{\smash{{\SetFigFontNFSS{10}{12.0}{\familydefault}{\mddefault}{\updefault} 1.6}}}}
\put(10886,-1722){\rotatebox{270.0}{\makebox(0,0)[b]{\smash{{\SetFigFontNFSS{10}{12.0}{\familydefault}{\mddefault}{\updefault}$\d N/\d\cos(\Delta \phi)\ [\unit{a.u.}]$}}}}}
\put(5329,-4008){\makebox(0,0)[rb]{\smash{{\SetFigFontNFSS{9}{10.8}{\familydefault}{\mddefault}{\updefault}$T=\unit[500]{MeV}$, recoil}}}}
\put(5329,-3809){\makebox(0,0)[rb]{\smash{{\SetFigFontNFSS{9}{10.8}{\familydefault}{\mddefault}{\updefault}$T=\unit[500]{MeV}$, medium}}}}
\put(5329,-3610){\makebox(0,0)[rb]{\smash{{\SetFigFontNFSS{9}{10.8}{\familydefault}{\mddefault}{\updefault}$T=\unit[200]{MeV}$, recoil}}}}
\put(5284,-3411){\makebox(0,0)[rb]{\smash{{\SetFigFontNFSS{9}{10.8}{\familydefault}{\mddefault}{\updefault}$T=\unit[200]{MeV}$, medium}}}}
\put(1776,-6221){\makebox(0,0)[rb]{\smash{{\SetFigFontNFSS{10}{12.0}{\familydefault}{\mddefault}{\updefault} 0.0001}}}}
\put(1776,-5630){\makebox(0,0)[rb]{\smash{{\SetFigFontNFSS{10}{12.0}{\familydefault}{\mddefault}{\updefault} 0.001}}}}
\put(1776,-5039){\makebox(0,0)[rb]{\smash{{\SetFigFontNFSS{10}{12.0}{\familydefault}{\mddefault}{\updefault} 0.01}}}}
\put(1776,-4449){\makebox(0,0)[rb]{\smash{{\SetFigFontNFSS{10}{12.0}{\familydefault}{\mddefault}{\updefault} 0.1}}}}
\put(1776,-3858){\makebox(0,0)[rb]{\smash{{\SetFigFontNFSS{10}{12.0}{\familydefault}{\mddefault}{\updefault} 1}}}}
\put(1868,-6375){\makebox(0,0)[b]{\smash{{\SetFigFontNFSS{10}{12.0}{\familydefault}{\mddefault}{\updefault} 0}}}}
\put(2285,-6375){\makebox(0,0)[b]{\smash{{\SetFigFontNFSS{10}{12.0}{\familydefault}{\mddefault}{\updefault} 1}}}}
\put(2702,-6375){\makebox(0,0)[b]{\smash{{\SetFigFontNFSS{10}{12.0}{\familydefault}{\mddefault}{\updefault} 2}}}}
\put(3120,-6375){\makebox(0,0)[b]{\smash{{\SetFigFontNFSS{10}{12.0}{\familydefault}{\mddefault}{\updefault} 3}}}}
\put(3537,-6375){\makebox(0,0)[b]{\smash{{\SetFigFontNFSS{10}{12.0}{\familydefault}{\mddefault}{\updefault} 4}}}}
\put(3954,-6375){\makebox(0,0)[b]{\smash{{\SetFigFontNFSS{10}{12.0}{\familydefault}{\mddefault}{\updefault} 5}}}}
\put(4371,-6375){\makebox(0,0)[b]{\smash{{\SetFigFontNFSS{10}{12.0}{\familydefault}{\mddefault}{\updefault} 6}}}}
\put(4788,-6375){\makebox(0,0)[b]{\smash{{\SetFigFontNFSS{10}{12.0}{\familydefault}{\mddefault}{\updefault} 7}}}}
\put(5206,-6375){\makebox(0,0)[b]{\smash{{\SetFigFontNFSS{10}{12.0}{\familydefault}{\mddefault}{\updefault} 8}}}}
\put(5623,-6375){\makebox(0,0)[b]{\smash{{\SetFigFontNFSS{10}{12.0}{\familydefault}{\mddefault}{\updefault} 9}}}}
\put(1131,-4667){\rotatebox{90.0}{\makebox(0,0)[b]{\smash{{\SetFigFontNFSS{10}{12.0}{\familydefault}{\mddefault}{\updefault}$\d N/\d\pt\ [\unit{a.u.}]$}}}}}
\put(3954,-6606){\makebox(0,0)[b]{\smash{{\SetFigFontNFSS{10}{12.0}{\familydefault}{\mddefault}{\updefault}$\pt\ [\unit{GeV}]$}}}}
\put(9692,-4008){\makebox(0,0)[rb]{\smash{{\SetFigFontNFSS{9}{10.8}{\familydefault}{\mddefault}{\updefault}$T=\unit[500]{MeV}$, recoil}}}}
\put(9692,-3809){\makebox(0,0)[rb]{\smash{{\SetFigFontNFSS{9}{10.8}{\familydefault}{\mddefault}{\updefault}$T=\unit[500]{MeV}$, medium}}}}
\put(9692,-3610){\makebox(0,0)[rb]{\smash{{\SetFigFontNFSS{9}{10.8}{\familydefault}{\mddefault}{\updefault}$T=\unit[200]{MeV}$, recoil}}}}
\put(9692,-3411){\makebox(0,0)[rb]{\smash{{\SetFigFontNFSS{9}{10.8}{\familydefault}{\mddefault}{\updefault}$T=\unit[200]{MeV}$, medium}}}}
\put(10423,-6221){\makebox(0,0)[lb]{\smash{{\SetFigFontNFSS{10}{12.0}{\familydefault}{\mddefault}{\updefault} 0}}}}
\put(10423,-5852){\makebox(0,0)[lb]{\smash{{\SetFigFontNFSS{10}{12.0}{\familydefault}{\mddefault}{\updefault} 0.2}}}}
\put(10423,-5482){\makebox(0,0)[lb]{\smash{{\SetFigFontNFSS{10}{12.0}{\familydefault}{\mddefault}{\updefault} 0.4}}}}
\put(10423,-5113){\makebox(0,0)[lb]{\smash{{\SetFigFontNFSS{10}{12.0}{\familydefault}{\mddefault}{\updefault} 0.6}}}}
\put(10423,-4744){\makebox(0,0)[lb]{\smash{{\SetFigFontNFSS{10}{12.0}{\familydefault}{\mddefault}{\updefault} 0.8}}}}
\put(10423,-4375){\makebox(0,0)[lb]{\smash{{\SetFigFontNFSS{10}{12.0}{\familydefault}{\mddefault}{\updefault} 1}}}}
\put(10423,-4006){\makebox(0,0)[lb]{\smash{{\SetFigFontNFSS{10}{12.0}{\familydefault}{\mddefault}{\updefault} 1.2}}}}
\put(10423,-3636){\makebox(0,0)[lb]{\smash{{\SetFigFontNFSS{10}{12.0}{\familydefault}{\mddefault}{\updefault} 1.4}}}}
\put(10884,-4667){\rotatebox{270.0}{\makebox(0,0)[b]{\smash{{\SetFigFontNFSS{10}{12.0}{\familydefault}{\mddefault}{\updefault}$\d N/\d\cos(\Delta \phi)\ [\unit{a.u}]$}}}}}
\put(6047,-6375){\makebox(0,0)[b]{\smash{{\SetFigFontNFSS{10}{12.0}{\familydefault}{\mddefault}{\updefault}-1}}}}
\put(6483,-6375){\makebox(0,0)[b]{\smash{{\SetFigFontNFSS{10}{12.0}{\familydefault}{\mddefault}{\updefault}-0.8}}}}
\put(6918,-6375){\makebox(0,0)[b]{\smash{{\SetFigFontNFSS{10}{12.0}{\familydefault}{\mddefault}{\updefault}-0.6}}}}
\put(7354,-6375){\makebox(0,0)[b]{\smash{{\SetFigFontNFSS{10}{12.0}{\familydefault}{\mddefault}{\updefault}-0.4}}}}
\put(7789,-6375){\makebox(0,0)[b]{\smash{{\SetFigFontNFSS{10}{12.0}{\familydefault}{\mddefault}{\updefault}-0.2}}}}
\put(8225,-6375){\makebox(0,0)[b]{\smash{{\SetFigFontNFSS{10}{12.0}{\familydefault}{\mddefault}{\updefault} 0}}}}
\put(8661,-6375){\makebox(0,0)[b]{\smash{{\SetFigFontNFSS{10}{12.0}{\familydefault}{\mddefault}{\updefault} 0.2}}}}
\put(9096,-6375){\makebox(0,0)[b]{\smash{{\SetFigFontNFSS{10}{12.0}{\familydefault}{\mddefault}{\updefault} 0.4}}}}
\put(9532,-6375){\makebox(0,0)[b]{\smash{{\SetFigFontNFSS{10}{12.0}{\familydefault}{\mddefault}{\updefault} 0.6}}}}
\put(9967,-6375){\makebox(0,0)[b]{\smash{{\SetFigFontNFSS{10}{12.0}{\familydefault}{\mddefault}{\updefault} 0.8}}}}
\put(10403,-6375){\makebox(0,0)[b]{\smash{{\SetFigFontNFSS{10}{12.0}{\familydefault}{\mddefault}{\updefault} 1}}}}
\put(8225,-6606){\makebox(0,0)[b]{\smash{{\SetFigFontNFSS{10}{12.0}{\familydefault}{\mddefault}{\updefault}$\cos(\Delta \phi)$}}}}
\put(5329,-1063){\makebox(0,0)[rb]{\smash{{\SetFigFontNFSS{9}{10.8}{\familydefault}{\mddefault}{\updefault}$T=\unit[500]{MeV}$, recoil}}}}
\put(5329,-864){\makebox(0,0)[rb]{\smash{{\SetFigFontNFSS{9}{10.8}{\familydefault}{\mddefault}{\updefault}$T=\unit[500]{MeV}$, medium}}}}
\put(5329,-665){\makebox(0,0)[rb]{\smash{{\SetFigFontNFSS{9}{10.8}{\familydefault}{\mddefault}{\updefault}$T=\unit[200]{MeV}$, recoil}}}}
\put(5329,-466){\makebox(0,0)[rb]{\smash{{\SetFigFontNFSS{9}{10.8}{\familydefault}{\mddefault}{\updefault}$T=\unit[200]{MeV}$, medium}}}}
\put(1776,-3276){\makebox(0,0)[rb]{\smash{{\SetFigFontNFSS{10}{12.0}{\familydefault}{\mddefault}{\updefault} 0.0001}}}}
\put(1776,-2685){\makebox(0,0)[rb]{\smash{{\SetFigFontNFSS{10}{12.0}{\familydefault}{\mddefault}{\updefault} 0.001}}}}
\put(1776,-2094){\makebox(0,0)[rb]{\smash{{\SetFigFontNFSS{10}{12.0}{\familydefault}{\mddefault}{\updefault} 0.01}}}}
\put(1776,-1504){\makebox(0,0)[rb]{\smash{{\SetFigFontNFSS{10}{12.0}{\familydefault}{\mddefault}{\updefault} 0.1}}}}
\put(1776,-913){\makebox(0,0)[rb]{\smash{{\SetFigFontNFSS{10}{12.0}{\familydefault}{\mddefault}{\updefault} 1}}}}
\put(1776,-322){\makebox(0,0)[rb]{\smash{{\SetFigFontNFSS{10}{12.0}{\familydefault}{\mddefault}{\updefault} 10}}}}
\put(1131,-1722){\rotatebox{90.0}{\makebox(0,0)[b]{\smash{{\SetFigFontNFSS{10}{12.0}{\familydefault}{\mddefault}{\updefault}$\d N/\d\pt\ [\unit{a.u.}]$}}}}}
\end{picture}%

%% file: 9-thrustmed-lqcd300.pstex_t
\begin{picture}(0,0)%
\includegraphics{9-thrustmed-lqcd300.pstex}%
\end{picture}%
\setlength{\unitlength}{3108sp}%
\begingroup\makeatletter\ifx\SetFigFontNFSS\undefined%
\gdef\SetFigFontNFSS#1#2#3#4#5{%
  \reset@font\fontsize{#1}{#2pt}%
  \fontfamily{#3}\fontseries{#4}\fontshape{#5}%
  \selectfont}%
\fi\endgroup%
\begin{picture}(9448,6534)(944,-6691)
\put(6285,-736){\makebox(0,0)[lb]{\smash{{\SetFigFontNFSS{10}{12.0}{\familydefault}{\mddefault}{\updefault}radiative}}}}
\put(6285,-1186){\makebox(0,0)[lb]{\smash{{\SetFigFontNFSS{10}{12.0}{\familydefault}{\mddefault}{\updefault}$E_\text{cut}=\unit[0]{GeV}$}}}}
\put(9588,-466){\makebox(0,0)[rb]{\smash{{\SetFigFontNFSS{9}{10.8}{\familydefault}{\mddefault}{\updefault}$T$ vacuum}}}}
\put(9588,-1461){\makebox(0,0)[rb]{\smash{{\SetFigFontNFSS{9}{10.8}{\familydefault}{\mddefault}{\updefault}$T_\text{maj}$ medium}}}}
\put(9588,-1262){\makebox(0,0)[rb]{\smash{{\SetFigFontNFSS{9}{10.8}{\familydefault}{\mddefault}{\updefault}$T_\text{maj}$ medium}}}}
\put(9588,-1063){\makebox(0,0)[rb]{\smash{{\SetFigFontNFSS{9}{10.8}{\familydefault}{\mddefault}{\updefault}$T$ medium}}}}
\put(9588,-864){\makebox(0,0)[rb]{\smash{{\SetFigFontNFSS{9}{10.8}{\familydefault}{\mddefault}{\updefault}$T_\text{maj}$ vacuum}}}}
\put(9588,-665){\makebox(0,0)[rb]{\smash{{\SetFigFontNFSS{9}{10.8}{\familydefault}{\mddefault}{\updefault}$T_\text{maj}$ vacuum}}}}
\put(6287,-3688){\makebox(0,0)[lb]{\smash{{\SetFigFontNFSS{10}{12.0}{\familydefault}{\mddefault}{\updefault}radiative}}}}
\put(6287,-4138){\makebox(0,0)[lb]{\smash{{\SetFigFontNFSS{10}{12.0}{\familydefault}{\mddefault}{\updefault}$E_\text{cut}=\unit[2]{GeV}$}}}}
\put(2022,-3688){\makebox(0,0)[lb]{\smash{{\SetFigFontNFSS{10}{12.0}{\familydefault}{\mddefault}{\updefault}collisional}}}}
\put(2022,-4138){\makebox(0,0)[lb]{\smash{{\SetFigFontNFSS{10}{12.0}{\familydefault}{\mddefault}{\updefault}$E_\text{cut}=\unit[2]{GeV}$}}}}
\put(2026,-736){\makebox(0,0)[lb]{\smash{{\SetFigFontNFSS{10}{12.0}{\familydefault}{\mddefault}{\updefault}collisional}}}}
\put(2026,-1186){\makebox(0,0)[lb]{\smash{{\SetFigFontNFSS{10}{12.0}{\familydefault}{\mddefault}{\updefault}$E_\text{cut}=\unit[0]{GeV}$}}}}
\put(5329,-1262){\makebox(0,0)[rb]{\smash{{\SetFigFontNFSS{9}{10.8}{\familydefault}{\mddefault}{\updefault}$T_\text{maj}$ medium}}}}
\put(5329,-1063){\makebox(0,0)[rb]{\smash{{\SetFigFontNFSS{9}{10.8}{\familydefault}{\mddefault}{\updefault}$T$ medium}}}}
\put(5329,-864){\makebox(0,0)[rb]{\smash{{\SetFigFontNFSS{9}{10.8}{\familydefault}{\mddefault}{\updefault}$T_\text{min}$ vacuum}}}}
\put(5329,-665){\makebox(0,0)[rb]{\smash{{\SetFigFontNFSS{9}{10.8}{\familydefault}{\mddefault}{\updefault}$T_\text{maj}$ vacuum}}}}
\put(5329,-466){\makebox(0,0)[rb]{\smash{{\SetFigFontNFSS{9}{10.8}{\familydefault}{\mddefault}{\updefault}$T$ vacuum}}}}
\put(5329,-1461){\makebox(0,0)[rb]{\smash{{\SetFigFontNFSS{9}{10.8}{\familydefault}{\mddefault}{\updefault}$T_\text{min}$ medium}}}}
\put(9590,-4413){\makebox(0,0)[rb]{\smash{{\SetFigFontNFSS{9}{10.8}{\familydefault}{\mddefault}{\updefault}$T_\text{maj}$ medium}}}}
\put(9590,-4214){\makebox(0,0)[rb]{\smash{{\SetFigFontNFSS{9}{10.8}{\familydefault}{\mddefault}{\updefault}$T_\text{maj}$ medium}}}}
\put(9590,-4015){\makebox(0,0)[rb]{\smash{{\SetFigFontNFSS{9}{10.8}{\familydefault}{\mddefault}{\updefault}$T$ medium}}}}
\put(9590,-3816){\makebox(0,0)[rb]{\smash{{\SetFigFontNFSS{9}{10.8}{\familydefault}{\mddefault}{\updefault}$T_\text{maj}$ vacuum}}}}
\put(9590,-3617){\makebox(0,0)[rb]{\smash{{\SetFigFontNFSS{9}{10.8}{\familydefault}{\mddefault}{\updefault}$T_\text{maj}$ vacuum}}}}
\put(9545,-3418){\makebox(0,0)[rb]{\smash{{\SetFigFontNFSS{9}{10.8}{\familydefault}{\mddefault}{\updefault}$T$ vacuum}}}}
\put(6037,-6382){\makebox(0,0)[b]{\smash{{\SetFigFontNFSS{10}{12.0}{\familydefault}{\mddefault}{\updefault} 0}}}}
\put(6463,-6382){\makebox(0,0)[b]{\smash{{\SetFigFontNFSS{10}{12.0}{\familydefault}{\mddefault}{\updefault} 0.1}}}}
\put(6890,-6382){\makebox(0,0)[b]{\smash{{\SetFigFontNFSS{10}{12.0}{\familydefault}{\mddefault}{\updefault} 0.2}}}}
\put(7316,-6382){\makebox(0,0)[b]{\smash{{\SetFigFontNFSS{10}{12.0}{\familydefault}{\mddefault}{\updefault} 0.3}}}}
\put(7743,-6382){\makebox(0,0)[b]{\smash{{\SetFigFontNFSS{10}{12.0}{\familydefault}{\mddefault}{\updefault} 0.4}}}}
\put(8169,-6382){\makebox(0,0)[b]{\smash{{\SetFigFontNFSS{10}{12.0}{\familydefault}{\mddefault}{\updefault} 0.5}}}}
\put(8595,-6382){\makebox(0,0)[b]{\smash{{\SetFigFontNFSS{10}{12.0}{\familydefault}{\mddefault}{\updefault} 0.6}}}}
\put(9022,-6382){\makebox(0,0)[b]{\smash{{\SetFigFontNFSS{10}{12.0}{\familydefault}{\mddefault}{\updefault} 0.7}}}}
\put(9448,-6382){\makebox(0,0)[b]{\smash{{\SetFigFontNFSS{10}{12.0}{\familydefault}{\mddefault}{\updefault} 0.8}}}}
\put(9875,-6382){\makebox(0,0)[b]{\smash{{\SetFigFontNFSS{10}{12.0}{\familydefault}{\mddefault}{\updefault} 0.9}}}}
\put(10301,-6382){\makebox(0,0)[b]{\smash{{\SetFigFontNFSS{10}{12.0}{\familydefault}{\mddefault}{\updefault} 1}}}}
\put(8169,-6613){\makebox(0,0)[b]{\smash{{\SetFigFontNFSS{10}{12.0}{\familydefault}{\mddefault}{\updefault}$T_\text{r}$}}}}
\put(5325,-3418){\makebox(0,0)[rb]{\smash{{\SetFigFontNFSS{9}{10.8}{\familydefault}{\mddefault}{\updefault}$T$ vacuum}}}}
\put(5325,-4413){\makebox(0,0)[rb]{\smash{{\SetFigFontNFSS{9}{10.8}{\familydefault}{\mddefault}{\updefault}$T_\text{maj}$ medium}}}}
\put(5325,-4214){\makebox(0,0)[rb]{\smash{{\SetFigFontNFSS{9}{10.8}{\familydefault}{\mddefault}{\updefault}$T_\text{maj}$ medium}}}}
\put(5325,-4015){\makebox(0,0)[rb]{\smash{{\SetFigFontNFSS{9}{10.8}{\familydefault}{\mddefault}{\updefault}$T$ medium}}}}
\put(5325,-3816){\makebox(0,0)[rb]{\smash{{\SetFigFontNFSS{9}{10.8}{\familydefault}{\mddefault}{\updefault}$T_\text{maj}$ vacuum}}}}
\put(5325,-3617){\makebox(0,0)[rb]{\smash{{\SetFigFontNFSS{9}{10.8}{\familydefault}{\mddefault}{\updefault}$T_\text{maj}$ vacuum}}}}
\put(1680,-6228){\makebox(0,0)[rb]{\smash{{\SetFigFontNFSS{10}{12.0}{\familydefault}{\mddefault}{\updefault} 0.1}}}}
\put(1680,-5490){\makebox(0,0)[rb]{\smash{{\SetFigFontNFSS{10}{12.0}{\familydefault}{\mddefault}{\updefault} 1}}}}
\put(1680,-4751){\makebox(0,0)[rb]{\smash{{\SetFigFontNFSS{10}{12.0}{\familydefault}{\mddefault}{\updefault} 10}}}}
\put(1680,-4012){\makebox(0,0)[rb]{\smash{{\SetFigFontNFSS{10}{12.0}{\familydefault}{\mddefault}{\updefault} 100}}}}
\put(1772,-6382){\makebox(0,0)[b]{\smash{{\SetFigFontNFSS{10}{12.0}{\familydefault}{\mddefault}{\updefault} 0}}}}
\put(2198,-6382){\makebox(0,0)[b]{\smash{{\SetFigFontNFSS{10}{12.0}{\familydefault}{\mddefault}{\updefault} 0.1}}}}
\put(2625,-6382){\makebox(0,0)[b]{\smash{{\SetFigFontNFSS{10}{12.0}{\familydefault}{\mddefault}{\updefault} 0.2}}}}
\put(3051,-6382){\makebox(0,0)[b]{\smash{{\SetFigFontNFSS{10}{12.0}{\familydefault}{\mddefault}{\updefault} 0.3}}}}
\put(3478,-6382){\makebox(0,0)[b]{\smash{{\SetFigFontNFSS{10}{12.0}{\familydefault}{\mddefault}{\updefault} 0.4}}}}
\put(3904,-6382){\makebox(0,0)[b]{\smash{{\SetFigFontNFSS{10}{12.0}{\familydefault}{\mddefault}{\updefault} 0.5}}}}
\put(4330,-6382){\makebox(0,0)[b]{\smash{{\SetFigFontNFSS{10}{12.0}{\familydefault}{\mddefault}{\updefault} 0.6}}}}
\put(4757,-6382){\makebox(0,0)[b]{\smash{{\SetFigFontNFSS{10}{12.0}{\familydefault}{\mddefault}{\updefault} 0.7}}}}
\put(5183,-6382){\makebox(0,0)[b]{\smash{{\SetFigFontNFSS{10}{12.0}{\familydefault}{\mddefault}{\updefault} 0.8}}}}
\put(5610,-6382){\makebox(0,0)[b]{\smash{{\SetFigFontNFSS{10}{12.0}{\familydefault}{\mddefault}{\updefault} 0.9}}}}
\put(1127,-4674){\rotatebox{90.0}{\makebox(0,0)[b]{\smash{{\SetFigFontNFSS{10}{12.0}{\familydefault}{\mddefault}{\updefault}$\d N/\d T_\text{r}$}}}}}
\put(3904,-6613){\makebox(0,0)[b]{\smash{{\SetFigFontNFSS{10}{12.0}{\familydefault}{\mddefault}{\updefault}$T_\text{r}$}}}}
\put(1684,-3276){\makebox(0,0)[rb]{\smash{{\SetFigFontNFSS{10}{12.0}{\familydefault}{\mddefault}{\updefault} 0.1}}}}
\put(1684,-2538){\makebox(0,0)[rb]{\smash{{\SetFigFontNFSS{10}{12.0}{\familydefault}{\mddefault}{\updefault} 1}}}}
\put(1684,-1799){\makebox(0,0)[rb]{\smash{{\SetFigFontNFSS{10}{12.0}{\familydefault}{\mddefault}{\updefault} 10}}}}
\put(1684,-1060){\makebox(0,0)[rb]{\smash{{\SetFigFontNFSS{10}{12.0}{\familydefault}{\mddefault}{\updefault} 100}}}}
\put(1684,-322){\makebox(0,0)[rb]{\smash{{\SetFigFontNFSS{10}{12.0}{\familydefault}{\mddefault}{\updefault} 1000}}}}
\put(1131,-1722){\rotatebox{90.0}{\makebox(0,0)[b]{\smash{{\SetFigFontNFSS{10}{12.0}{\familydefault}{\mddefault}{\updefault}$\d N/\d T_\text{r}$}}}}}
\end{picture}%

%% file: 8-njetmed-lqcd300.pstex_t
\begin{picture}(0,0)%
\includegraphics{8-njetmed-lqcd300.pstex}%
\end{picture}%
\setlength{\unitlength}{3108sp}%
\begingroup\makeatletter\ifx\SetFigFontNFSS\undefined%
\gdef\SetFigFontNFSS#1#2#3#4#5{%
  \reset@font\fontsize{#1}{#2pt}%
  \fontfamily{#3}\fontseries{#4}\fontshape{#5}%
  \selectfont}%
\fi\endgroup%
\begin{picture}(9904,6743)(876,-6934)
\put(2071,-1186){\makebox(0,0)[lb]{\smash{{\SetFigFontNFSS{10}{12.0}{\familydefault}{\mddefault}{\updefault}collisional}}}}
\put(2026,-1546){\makebox(0,0)[lb]{\smash{{\SetFigFontNFSS{10}{12.0}{\familydefault}{\mddefault}{\updefault}$E_\text{cut}=\unit[0]{GeV}$}}}}
\put(2071,-4291){\makebox(0,0)[lb]{\smash{{\SetFigFontNFSS{10}{12.0}{\familydefault}{\mddefault}{\updefault}collisional}}}}
\put(2026,-4696){\makebox(0,0)[lb]{\smash{{\SetFigFontNFSS{10}{12.0}{\familydefault}{\mddefault}{\updefault}$E_\text{cut}=\unit[2]{GeV}$}}}}
\put(6661,-4696){\makebox(0,0)[lb]{\smash{{\SetFigFontNFSS{10}{12.0}{\familydefault}{\mddefault}{\updefault}$E_\text{cut}=\unit[2]{GeV}$}}}}
\put(6661,-4291){\makebox(0,0)[lb]{\smash{{\SetFigFontNFSS{10}{12.0}{\familydefault}{\mddefault}{\updefault}radiative}}}}
\put(6661,-1546){\makebox(0,0)[lb]{\smash{{\SetFigFontNFSS{10}{12.0}{\familydefault}{\mddefault}{\updefault}$E_\text{cut}=\unit[0]{GeV}$}}}}
\put(6661,-1186){\makebox(0,0)[lb]{\smash{{\SetFigFontNFSS{10}{12.0}{\familydefault}{\mddefault}{\updefault}radiative}}}}
\put(5535,-556){\makebox(0,0)[rb]{\smash{{\SetFigFontNFSS{8}{9.6}{\familydefault}{\mddefault}{\updefault}medium, hadronis.: all}}}}
\put(5535,-3661){\makebox(0,0)[rb]{\smash{{\SetFigFontNFSS{8}{9.6}{\familydefault}{\mddefault}{\updefault}medium, hadronis.: all}}}}
\put(2835,-716){\makebox(0,0)[rb]{\smash{{\SetFigFontNFSS{8}{9.6}{\familydefault}{\mddefault}{\updefault}vacuum, 3 jets}}}}
\put(2835,-552){\makebox(0,0)[rb]{\smash{{\SetFigFontNFSS{8}{9.6}{\familydefault}{\mddefault}{\updefault}vacuum, 2 jets}}}}
\put(2835,-388){\makebox(0,0)[rb]{\smash{{\SetFigFontNFSS{8}{9.6}{\familydefault}{\mddefault}{\updefault}vacuum, 1 jet}}}}
\put(5535,-385){\makebox(0,0)[rb]{\smash{{\SetFigFontNFSS{8}{9.6}{\familydefault}{\mddefault}{\updefault}medium, hadronis.: cascade}}}}
\put(2790,-3493){\makebox(0,0)[rb]{\smash{{\SetFigFontNFSS{8}{9.6}{\familydefault}{\mddefault}{\updefault}vacuum, 1 jet}}}}
\put(2790,-3657){\makebox(0,0)[rb]{\smash{{\SetFigFontNFSS{8}{9.6}{\familydefault}{\mddefault}{\updefault}vacuum, 2 jets}}}}
\put(2790,-3821){\makebox(0,0)[rb]{\smash{{\SetFigFontNFSS{8}{9.6}{\familydefault}{\mddefault}{\updefault}vacuum, 3 jets}}}}
\put(5535,-3490){\makebox(0,0)[rb]{\smash{{\SetFigFontNFSS{8}{9.6}{\familydefault}{\mddefault}{\updefault}medium, hadronis.: cascade}}}}
\put(1505,-6632){\makebox(0,0)[b]{\smash{{\SetFigFontNFSS{10}{12.0}{\familydefault}{\mddefault}{\updefault}-4}}}}
\put(2079,-6632){\makebox(0,0)[b]{\smash{{\SetFigFontNFSS{10}{12.0}{\familydefault}{\mddefault}{\updefault}-3.5}}}}
\put(2654,-6632){\makebox(0,0)[b]{\smash{{\SetFigFontNFSS{10}{12.0}{\familydefault}{\mddefault}{\updefault}-3}}}}
\put(3228,-6632){\makebox(0,0)[b]{\smash{{\SetFigFontNFSS{10}{12.0}{\familydefault}{\mddefault}{\updefault}-2.5}}}}
\put(3802,-6632){\makebox(0,0)[b]{\smash{{\SetFigFontNFSS{10}{12.0}{\familydefault}{\mddefault}{\updefault}-2}}}}
\put(4376,-6632){\makebox(0,0)[b]{\smash{{\SetFigFontNFSS{10}{12.0}{\familydefault}{\mddefault}{\updefault}-1.5}}}}
\put(4951,-6632){\makebox(0,0)[b]{\smash{{\SetFigFontNFSS{10}{12.0}{\familydefault}{\mddefault}{\updefault}-1}}}}
\put(5525,-6632){\makebox(0,0)[b]{\smash{{\SetFigFontNFSS{10}{12.0}{\familydefault}{\mddefault}{\updefault}-0.5}}}}
\put(3802,-6856){\makebox(0,0)[b]{\smash{{\SetFigFontNFSS{10}{12.0}{\familydefault}{\mddefault}{\updefault}$\log_{10}(y_\text{cut})$}}}}
\put(10125,-3493){\makebox(0,0)[rb]{\smash{{\SetFigFontNFSS{8}{9.6}{\familydefault}{\mddefault}{\updefault}vacuum, 1 jet}}}}
\put(10125,-3985){\makebox(0,0)[rb]{\smash{{\SetFigFontNFSS{8}{9.6}{\familydefault}{\mddefault}{\updefault}medium}}}}
\put(10125,-3821){\makebox(0,0)[rb]{\smash{{\SetFigFontNFSS{8}{9.6}{\familydefault}{\mddefault}{\updefault}vacuum, 3 jets}}}}
\put(10125,-3657){\makebox(0,0)[rb]{\smash{{\SetFigFontNFSS{8}{9.6}{\familydefault}{\mddefault}{\updefault}vacuum, 2 jets}}}}
\put(6669,-6632){\makebox(0,0)[b]{\smash{{\SetFigFontNFSS{10}{12.0}{\familydefault}{\mddefault}{\updefault}-3.5}}}}
\put(7244,-6632){\makebox(0,0)[b]{\smash{{\SetFigFontNFSS{10}{12.0}{\familydefault}{\mddefault}{\updefault}-3}}}}
\put(7818,-6632){\makebox(0,0)[b]{\smash{{\SetFigFontNFSS{10}{12.0}{\familydefault}{\mddefault}{\updefault}-2.5}}}}
\put(8392,-6632){\makebox(0,0)[b]{\smash{{\SetFigFontNFSS{10}{12.0}{\familydefault}{\mddefault}{\updefault}-2}}}}
\put(8966,-6632){\makebox(0,0)[b]{\smash{{\SetFigFontNFSS{10}{12.0}{\familydefault}{\mddefault}{\updefault}-1.5}}}}
\put(9541,-6632){\makebox(0,0)[b]{\smash{{\SetFigFontNFSS{10}{12.0}{\familydefault}{\mddefault}{\updefault}-1}}}}
\put(10115,-6632){\makebox(0,0)[b]{\smash{{\SetFigFontNFSS{10}{12.0}{\familydefault}{\mddefault}{\updefault}-0.5}}}}
\put(10689,-6632){\makebox(0,0)[b]{\smash{{\SetFigFontNFSS{10}{12.0}{\familydefault}{\mddefault}{\updefault} 0}}}}
\put(8392,-6856){\makebox(0,0)[b]{\smash{{\SetFigFontNFSS{10}{12.0}{\familydefault}{\mddefault}{\updefault}$\log_{10}(y_\text{cut})$}}}}
\put(6095,-6632){\makebox(0,0)[b]{\smash{{\SetFigFontNFSS{10}{12.0}{\familydefault}{\mddefault}{\updefault}-4}}}}
\put(10125,-388){\makebox(0,0)[rb]{\smash{{\SetFigFontNFSS{8}{9.6}{\familydefault}{\mddefault}{\updefault}vacuum, 1 jet}}}}
\put(10125,-880){\makebox(0,0)[rb]{\smash{{\SetFigFontNFSS{8}{9.6}{\familydefault}{\mddefault}{\updefault}medium}}}}
\put(10125,-716){\makebox(0,0)[rb]{\smash{{\SetFigFontNFSS{8}{9.6}{\familydefault}{\mddefault}{\updefault}vacuum, 3 jets}}}}
\put(10125,-552){\makebox(0,0)[rb]{\smash{{\SetFigFontNFSS{8}{9.6}{\familydefault}{\mddefault}{\updefault}vacuum, 2 jets}}}}
\put(1434,-3304){\makebox(0,0)[rb]{\smash{{\SetFigFontNFSS{10}{12.0}{\familydefault}{\mddefault}{\updefault} 0}}}}
\put(1434,-2836){\makebox(0,0)[rb]{\smash{{\SetFigFontNFSS{10}{12.0}{\familydefault}{\mddefault}{\updefault} 0.2}}}}
\put(1048,-1754){\rotatebox{90.0}{\makebox(0,0)[b]{\smash{{\SetFigFontNFSS{10}{12.0}{\familydefault}{\mddefault}{\updefault}n-jet fraction}}}}}
\put(1434,-2368){\makebox(0,0)[rb]{\smash{{\SetFigFontNFSS{10}{12.0}{\familydefault}{\mddefault}{\updefault} 0.4}}}}
\put(1434,-1900){\makebox(0,0)[rb]{\smash{{\SetFigFontNFSS{10}{12.0}{\familydefault}{\mddefault}{\updefault} 0.6}}}}
\put(1434,-1432){\makebox(0,0)[rb]{\smash{{\SetFigFontNFSS{10}{12.0}{\familydefault}{\mddefault}{\updefault} 0.8}}}}
\put(1434,-964){\makebox(0,0)[rb]{\smash{{\SetFigFontNFSS{10}{12.0}{\familydefault}{\mddefault}{\updefault} 1}}}}
\put(1434,-496){\makebox(0,0)[rb]{\smash{{\SetFigFontNFSS{10}{12.0}{\familydefault}{\mddefault}{\updefault} 1.2}}}}
\put(1048,-4859){\rotatebox{90.0}{\makebox(0,0)[b]{\smash{{\SetFigFontNFSS{10}{12.0}{\familydefault}{\mddefault}{\updefault}n-jet fraction}}}}}
\put(1434,-3601){\makebox(0,0)[rb]{\smash{{\SetFigFontNFSS{10}{12.0}{\familydefault}{\mddefault}{\updefault} 1.2}}}}
\put(1434,-4069){\makebox(0,0)[rb]{\smash{{\SetFigFontNFSS{10}{12.0}{\familydefault}{\mddefault}{\updefault} 1}}}}
\put(1434,-4537){\makebox(0,0)[rb]{\smash{{\SetFigFontNFSS{10}{12.0}{\familydefault}{\mddefault}{\updefault} 0.8}}}}
\put(1434,-5005){\makebox(0,0)[rb]{\smash{{\SetFigFontNFSS{10}{12.0}{\familydefault}{\mddefault}{\updefault} 0.6}}}}
\put(1434,-5473){\makebox(0,0)[rb]{\smash{{\SetFigFontNFSS{10}{12.0}{\familydefault}{\mddefault}{\updefault} 0.4}}}}
\put(1434,-5941){\makebox(0,0)[rb]{\smash{{\SetFigFontNFSS{10}{12.0}{\familydefault}{\mddefault}{\updefault} 0.2}}}}
\put(1434,-6409){\makebox(0,0)[rb]{\smash{{\SetFigFontNFSS{10}{12.0}{\familydefault}{\mddefault}{\updefault} 0}}}}
\end{picture}%

%% file: 7-ximed-lqcd300.pstex_t
\begin{picture}(0,0)%
\includegraphics{7-ximed-lqcd300.pstex}%
\end{picture}%
\setlength{\unitlength}{3108sp}%
\begingroup\makeatletter\ifx\SetFigFontNFSS\undefined%
\gdef\SetFigFontNFSS#1#2#3#4#5{%
  \reset@font\fontsize{#1}{#2pt}%
  \fontfamily{#3}\fontseries{#4}\fontshape{#5}%
  \selectfont}%
\fi\endgroup%
\begin{picture}(9618,3573)(960,-3730)
\put(6473,-961){\makebox(0,0)[lb]{\smash{{\SetFigFontNFSS{10}{12.0}{\familydefault}{\mddefault}{\updefault}hadron level}}}}
\put(2026,-961){\makebox(0,0)[lb]{\smash{{\SetFigFontNFSS{10}{12.0}{\familydefault}{\mddefault}{\updefault}parton level}}}}
\put(5329,-1063){\makebox(0,0)[rb]{\smash{{\SetFigFontNFSS{9}{10.8}{\familydefault}{\mddefault}{\updefault}radiative}}}}
\put(5329,-864){\makebox(0,0)[rb]{\smash{{\SetFigFontNFSS{9}{10.8}{\familydefault}{\mddefault}{\updefault}collisional: all}}}}
\put(5329,-665){\makebox(0,0)[rb]{\smash{{\SetFigFontNFSS{9}{10.8}{\familydefault}{\mddefault}{\updefault}collisional: cascade}}}}
\put(9776,-1063){\makebox(0,0)[rb]{\smash{{\SetFigFontNFSS{9}{10.8}{\familydefault}{\mddefault}{\updefault}radiative}}}}
\put(9776,-864){\makebox(0,0)[rb]{\smash{{\SetFigFontNFSS{9}{10.8}{\familydefault}{\mddefault}{\updefault}collisional: all}}}}
\put(9776,-665){\makebox(0,0)[rb]{\smash{{\SetFigFontNFSS{9}{10.8}{\familydefault}{\mddefault}{\updefault}collisional: cascade}}}}
\put(9776,-466){\makebox(0,0)[rb]{\smash{{\SetFigFontNFSS{9}{10.8}{\familydefault}{\mddefault}{\updefault}vacuum}}}}
\put(6039,-3430){\makebox(0,0)[b]{\smash{{\SetFigFontNFSS{10}{12.0}{\familydefault}{\mddefault}{\updefault} 0}}}}
\put(6595,-3430){\makebox(0,0)[b]{\smash{{\SetFigFontNFSS{10}{12.0}{\familydefault}{\mddefault}{\updefault} 1}}}}
\put(7151,-3430){\makebox(0,0)[b]{\smash{{\SetFigFontNFSS{10}{12.0}{\familydefault}{\mddefault}{\updefault} 2}}}}
\put(7707,-3430){\makebox(0,0)[b]{\smash{{\SetFigFontNFSS{10}{12.0}{\familydefault}{\mddefault}{\updefault} 3}}}}
\put(8263,-3430){\makebox(0,0)[b]{\smash{{\SetFigFontNFSS{10}{12.0}{\familydefault}{\mddefault}{\updefault} 4}}}}
\put(8819,-3430){\makebox(0,0)[b]{\smash{{\SetFigFontNFSS{10}{12.0}{\familydefault}{\mddefault}{\updefault} 5}}}}
\put(9375,-3430){\makebox(0,0)[b]{\smash{{\SetFigFontNFSS{10}{12.0}{\familydefault}{\mddefault}{\updefault} 6}}}}
\put(9931,-3430){\makebox(0,0)[b]{\smash{{\SetFigFontNFSS{10}{12.0}{\familydefault}{\mddefault}{\updefault} 7}}}}
\put(10487,-3430){\makebox(0,0)[b]{\smash{{\SetFigFontNFSS{10}{12.0}{\familydefault}{\mddefault}{\updefault} 8}}}}
\put(8263,-3661){\makebox(0,0)[b]{\smash{{\SetFigFontNFSS{10}{12.0}{\familydefault}{\mddefault}{\updefault}$\xi$}}}}
\put(1500,-3276){\makebox(0,0)[rb]{\smash{{\SetFigFontNFSS{10}{12.0}{\familydefault}{\mddefault}{\updefault} 0}}}}
\put(1500,-2948){\makebox(0,0)[rb]{\smash{{\SetFigFontNFSS{10}{12.0}{\familydefault}{\mddefault}{\updefault} 2}}}}
\put(1500,-2620){\makebox(0,0)[rb]{\smash{{\SetFigFontNFSS{10}{12.0}{\familydefault}{\mddefault}{\updefault} 4}}}}
\put(1500,-2291){\makebox(0,0)[rb]{\smash{{\SetFigFontNFSS{10}{12.0}{\familydefault}{\mddefault}{\updefault} 6}}}}
\put(1500,-1963){\makebox(0,0)[rb]{\smash{{\SetFigFontNFSS{10}{12.0}{\familydefault}{\mddefault}{\updefault} 8}}}}
\put(1500,-1635){\makebox(0,0)[rb]{\smash{{\SetFigFontNFSS{10}{12.0}{\familydefault}{\mddefault}{\updefault} 10}}}}
\put(1500,-1307){\makebox(0,0)[rb]{\smash{{\SetFigFontNFSS{10}{12.0}{\familydefault}{\mddefault}{\updefault} 12}}}}
\put(1500,-978){\makebox(0,0)[rb]{\smash{{\SetFigFontNFSS{10}{12.0}{\familydefault}{\mddefault}{\updefault} 14}}}}
\put(1500,-650){\makebox(0,0)[rb]{\smash{{\SetFigFontNFSS{10}{12.0}{\familydefault}{\mddefault}{\updefault} 16}}}}
\put(1500,-322){\makebox(0,0)[rb]{\smash{{\SetFigFontNFSS{10}{12.0}{\familydefault}{\mddefault}{\updefault} 18}}}}
\put(1592,-3430){\makebox(0,0)[b]{\smash{{\SetFigFontNFSS{10}{12.0}{\familydefault}{\mddefault}{\updefault} 0}}}}
\put(2148,-3430){\makebox(0,0)[b]{\smash{{\SetFigFontNFSS{10}{12.0}{\familydefault}{\mddefault}{\updefault} 1}}}}
\put(2704,-3430){\makebox(0,0)[b]{\smash{{\SetFigFontNFSS{10}{12.0}{\familydefault}{\mddefault}{\updefault} 2}}}}
\put(3260,-3430){\makebox(0,0)[b]{\smash{{\SetFigFontNFSS{10}{12.0}{\familydefault}{\mddefault}{\updefault} 3}}}}
\put(3816,-3430){\makebox(0,0)[b]{\smash{{\SetFigFontNFSS{10}{12.0}{\familydefault}{\mddefault}{\updefault} 4}}}}
\put(4372,-3430){\makebox(0,0)[b]{\smash{{\SetFigFontNFSS{10}{12.0}{\familydefault}{\mddefault}{\updefault} 5}}}}
\put(4928,-3430){\makebox(0,0)[b]{\smash{{\SetFigFontNFSS{10}{12.0}{\familydefault}{\mddefault}{\updefault} 6}}}}
\put(5484,-3430){\makebox(0,0)[b]{\smash{{\SetFigFontNFSS{10}{12.0}{\familydefault}{\mddefault}{\updefault} 7}}}}
\put(1131,-1722){\rotatebox{90.0}{\makebox(0,0)[b]{\smash{{\SetFigFontNFSS{10}{12.0}{\familydefault}{\mddefault}{\updefault}$\d N/\d \xi$}}}}}
\put(3816,-3661){\makebox(0,0)[b]{\smash{{\SetFigFontNFSS{10}{12.0}{\familydefault}{\mddefault}{\updefault}$\xi$}}}}
\put(5329,-466){\makebox(0,0)[rb]{\smash{{\SetFigFontNFSS{9}{10.8}{\familydefault}{\mddefault}{\updefault}vacuum}}}}
\end{picture}%

%% file: korinna18.bbl
\begin{thebibliography}{99}
%
\bibitem{Adcox:2004mh}
  K.~Adcox {\it et al.}  [PHENIX Collaboration],
  Nucl.\ Phys.\ A {\bf 757} (2005) 184.
%
\bibitem{Back:2004je}
  B.~B.~Back {\it et al.} [PHOBOS Collaboration],
  Nucl.\ Phys.\ A {\bf 757} (2005) 28.
%
\bibitem{Arsene:2004fa}
  I.~Arsene {\it et al.}  [BRAHMS Collaboration],
  Nucl.\ Phys.\ A {\bf 757} (2005) 1.
%
\bibitem{Adams:2005dq}
  J.~Adams {\it et al.}  [STAR Collaboration],
  Nucl.\ Phys.\ A {\bf 757} (2005) 102.

\bibitem{Carminati:2004fp}
  F.~Carminati {\it et al.}  [ALICE Collaboration],
  J.\ Phys.\ G {\bf 30} (2004) 1517.
  
\bibitem{Alessandro:2006yt}
  B.~Alessandro {\it et al.}  [ALICE Collaboration],
  J.\ Phys.\ G {\bf 32} (2006) 1295.

\bibitem{D'Enterria:2007xr}
  D.~d'Enterria {\it et al.} [CMS Collaboration],
  J.\ Phys.\ G {\bf 34} (2007) 2307.


\bibitem{Bjorken:1982tu}
  J.~D.~Bjorken, {\it preprint Fermilab-pub-82-059-thy}.

\bibitem{Thoma:1990fm}
  M.~H.~Thoma and M.~Gyulassy,
  Nucl.\ Phys.\  B {\bf 351} (1991) 491.

\bibitem{Braaten:1991we}
  E.~Braaten and M.~H.~Thoma,
  Phys.\ Rev.\  D {\bf 44} (1991) 2625.

\bibitem{Djordjevic:2006tw}
  M.~Djordjevic,
  Phys.\ Rev.\  C {\bf 74} (2006) 064907
  [arXiv:nucl-th/0603066].

\bibitem{Adil:2006ei}
  A.~Adil, M.~Gyulassy, W.~A.~Horowitz and S.~Wicks,
  Phys.\ Rev.\  C {\bf 75} (2007) 044906
  [arXiv:nucl-th/0606010].

\bibitem{Zakharov:2007pj}
  B.~G.~Zakharov,
  JETP Lett.\  {\bf 86} (2007) 444
  [arXiv:0708.0816 [hep-ph]].

\bibitem{Peigne:2008nd}
  S.~Peign\'e and A.~Peshier,
  arXiv:0802.4364 [hep-ph].

\bibitem{Zapp:2005kt}
  K.~Zapp, G.~Ingelman, J.~Rathsman and J.~Stachel,
  Phys.\ Lett.\  B {\bf 637} (2006) 179
  [arXiv:hep-ph/0512300].

\bibitem{Domdey:2008gp}
  S.~Domdey, G.~Ingelman, H.~J.~Pirner, J.~Rathsman, J.~Stachel and K.~Zapp,
  arXiv:0802.3282 [hep-ph].

%
\bibitem{Gyulassy:1993hr}
M.~Gyulassy and X.~N.~Wang,
Nucl.\ Phys.\ B {\bf 420} (1994) 583.
%
\bibitem{Baier:1996sk}
R.~Baier, Y.~L.~Dokshitzer, A.~H.~Mueller, S.~Peign\'e and D.~Schiff,
Nucl.\ Phys.\ B {\bf 484} (1997) 265.
%
\bibitem{Zakharov:1997uu}
B.~G.~Zakharov,
JETP Lett.\  {\bf 65} (1997) 615.
%
\bibitem{Wiedemann:2000za}
U.~A.~Wiedemann,
Nucl.\ Phys.\ B {\bf 588} (2000) 303.
%
\bibitem{Gyulassy:2000er}
M.~Gyulassy, P.~Levai and I.~Vitev,
Nucl.\ Phys.\ B {\bf 594} (2001) 371.
%
\bibitem{Wang:2001if}
X.~N.~Wang and X.~F.~Guo,
Nucl.\ Phys.\ A {\bf 696} (2001) 788.


\bibitem{Sjostrand:2006za}
  T.~Sjostrand, S.~Mrenna and P.~Skands,
  JHEP {\bf 0605} (2006) 026
  [arXiv:hep-ph/0603175].

\bibitem{Marchesini:1991ch}
  G.~Marchesini, B.~R.~Webber, G.~Abbiendi, I.~G.~Knowles, M.~H.~Seymour and L.~Stanco,
  Comput.\ Phys.\ Commun.\  {\bf 67} (1992) 465.

\bibitem{Gleisberg:2003xi}
  T.~Gleisberg, S.~Hoche, F.~Krauss, A.~Schalicke, S.~Schumann and J.~C.~Winter,
  JHEP {\bf 0402} (2004) 056
  [arXiv:hep-ph/0311263].


\bibitem{Baier:2001yt}
  R.~Baier, Y.~L.~Dokshitzer, A.~H.~Mueller and D.~Schiff,
  JHEP {\bf 0109} (2001) 033
  [arXiv:hep-ph/0106347].

\bibitem{Salgado:2003gb}
  C.~A.~Salgado and U.~A.~Wiedemann,
  Phys.\ Rev.\  D {\bf 68} (2003) 014008
  [arXiv:hep-ph/0302184].

\bibitem{Armesto:2007dt}
  N.~Armesto, L.~Cunqueiro, C.~A.~Salgado and W.~C.~Xiang,
  JHEP {\bf 0802} (2008) 048
  [arXiv:0710.3073 [hep-ph]].
  
\bibitem{Majumder:2004br}
  A.~Majumder and X.~N.~Wang,
  Phys.\ Rev.\  D {\bf 72} (2005) 034007
  [arXiv:hep-ph/0411174].

\bibitem{Salgado:2003rv}
  C.~A.~Salgado and U.~A.~Wiedemann,
  Phys.\ Rev.\ Lett.\  {\bf 93} (2004) 042301
  [arXiv:hep-ph/0310079].

\bibitem{Polosa:2006hb}
  A.~D.~Polosa and C.~A.~Salgado,
  Phys.\ Rev.\  C {\bf 75} (2007) 041901
  [arXiv:hep-ph/0607295].


\bibitem{Armesto:2004pt}
  N.~Armesto, C.~A.~Salgado and U.~A.~Wiedemann,
  Phys.\ Rev.\ Lett.\  {\bf 93} (2004) 242301
  [arXiv:hep-ph/0405301].
  
\bibitem{Armesto:2004vz}
  N.~Armesto, C.~A.~Salgado and U.~A.~Wiedemann,
  Phys.\ Rev.\  C {\bf 72} (2005) 064910
  [arXiv:hep-ph/0411341].

\bibitem{Borghini:2005em}
  N.~Borghini and U.~A.~Wiedemann,
  arXiv:hep-ph/0506218.

\bibitem{Sapeta:2007ad}
  S.~Sapeta and U.~A.~Wiedemann,
  arXiv:0707.3494 [hep-ph].


\bibitem{Lokhtin:2005px}
  I.~P.~Lokhtin and A.~M.~Snigirev,
  Eur.\ Phys.\ J.\  C {\bf 45} (2006) 211
  [arXiv:hep-ph/0506189].
    
\bibitem{Wang:1991hta}
  X.~N.~Wang and M.~Gyulassy,
  Phys.\ Rev.\  D {\bf 44} (1991) 3501.

\bibitem{Dainese:2004te}
  A.~Dainese, C.~Loizides and G.~Paic,
  Eur.\ Phys.\ J.\  C {\bf 38} (2005) 461
  [arXiv:hep-ph/0406201].

\bibitem{Armesto:2008qh}
  N.~Armesto, L.~Cunqueiro and C.~A.~Salgado,
  arXiv:0809.4433 [hep-ph].

\bibitem{Renk:2008pp}
  T.~Renk,
  Phys.\ Rev.\  C {\bf 78} (2008) 034908
  [arXiv:0806.0305 [hep-ph]].

\bibitem{lund}
  B.~Andersson, G.~Gustafson, G.~Ingelman and T.~Sjostrand,
  Phys.\ Rept.\  {\bf 97} (1983) 31.

\bibitem{Heister:2003aj}
  A.~Heister {\it et al.}  [ALEPH Collaboration],
  Eur.\ Phys.\ J.\  C {\bf 35} (2004) 457.

\bibitem{Korchemsky:1995zm}
  G.~P.~Korchemsky and G.~Sterman,
  arXiv:hep-ph/9505391.

\bibitem{Dokshitzer:1997ew}
  Y.~L.~Dokshitzer and B.~R.~Webber,
  Phys.\ Lett.\  B {\bf 404} (1997) 321
  [arXiv:hep-ph/9704298].

\bibitem{Catani:1991hj}
  S.~Catani, Y.~L.~Dokshitzer, M.~Olsson, G.~Turnock and B.~R.~Webber,
  Phys.\ Lett.\  B {\bf 269} (1991) 432.

\bibitem{Finkelstein:1972qm}
  J.~Finkelstein and R.~D.~Peccei,
  Phys.\ Rev.\  D {\bf 6}, 2606 (1972).

\bibitem{Krzywicki:1973qh}
  A.~Krzywicki and B.~Petersson,
  Phys.\ Rev.\  D {\bf 6}, 924 (1973).
  
\bibitem{Field:1977fa}
  R.~D.~Field and R.~P.~Feynman,
  Nucl.\ Phys.\  B {\bf 136}, 1 (1978).

\bibitem{Molnar:2003ff}
  D.~Molnar and S.~A.~Voloshin,
  Phys.\ Rev.\ Lett.\  {\bf 91}, 092301 (2003)
  [arXiv:nucl-th/0302014].

\bibitem{Fries:2003vb}
  R.~J.~Fries, B.~M\"uller, C.~Nonaka and S.~A.~Bass,
  Phys.\ Rev.\ Lett.\  {\bf 90}, 202303 (2003)
  [arXiv:nucl-th/0301087].

\bibitem{Greco:2003xt}
  V.~Greco, C.~M.~Ko and P.~Levai,
  Phys.\ Rev.\ Lett.\  {\bf 90}, 202302 (2003)
  [arXiv:nucl-th/0301093].

\bibitem{Hwa:2002tu}
  R.~C.~Hwa and C.~B.~Yang,
  Phys.\ Rev.\  C {\bf 67}, 034902 (2003)
  [arXiv:nucl-th/0211010].

\bibitem{Fries:2003kq}
  R.~J.~Fries, B.~M\"uller, C.~Nonaka and S.~A.~Bass,
  Phys.\ Rev.\  C {\bf 68}, 044902 (2003)
  [arXiv:nucl-th/0306027].
 
\bibitem{Liu:2008zb}
  W.~Liu and R.~J.~Fries,
  arXiv:0801.0453 [nucl-th].
  
\bibitem{Combridge:1978kx}
  B.~L.~Combridge,
  Nucl.\ Phys.\  B {\bf 151} (1979) 429.
  
\bibitem{Zapp:2007zs}
  K.~Zapp, G.~Ingelman, J.~Rathsman and J.~Stachel,
  Int.\ J.\ Mod.\ Phys.\  E {\bf 16}, 2072 (2007)
  [arXiv:hep-ph/0702201].


\bibitem{Kolb:2001qz}
  P.~F.~Kolb, U.~W.~Heinz, P.~Huovinen, K.~J.~Eskola and K.~Tuominen,
  Nucl.\ Phys.\  A {\bf 696} (2001) 197
  [arXiv:hep-ph/0103234].

\bibitem{Song:2007ux}
  H.~Song and U.~W.~Heinz,
  arXiv:0712.3715 [nucl-th].

\bibitem{Romatschke:2007mq}
  P.~Romatschke and U.~Romatschke,
  Phys.\ Rev.\ Lett.\  {\bf 99} (2007) 172301
  [arXiv:0706.1522 [nucl-th]].
  
\bibitem{Renk:2006sx}
  T.~Renk, J.~Ruppert, C.~Nonaka and S.~A.~Bass,
  Phys.\ Rev.\  C {\bf 75} (2007) 031902
  [arXiv:nucl-th/0611027].
  
\bibitem{Cacciari:2005hq}
  M.~Cacciari and G.~P.~Salam,
  Phys.\ Lett.\  B {\bf 641}, 57 (2006)
  [arXiv:hep-ph/0512210].
  

\end{thebibliography}
